\documentclass[floatfix,aps,twocolumn,nofootinbib,superscriptaddress,preprintnumbers,prd,10pt]{revtex4-2}
\pdfoutput=1
\usepackage[title]{appendix}
\usepackage[colorlinks=true,breaklinks=true]{hyperref}
\usepackage[normalem]{ulem}
\usepackage[utf8]{inputenc}
\hypersetup{allcolors=[rgb]{0.0 0.0 0.6},linkcolor=[rgb]{0.75 0.05 0.05}}
\usepackage{amsmath,amssymb}
\usepackage{epsfig}  
\usepackage{graphicx}   
\usepackage{slashed}       
\usepackage{url}
\usepackage[dvipsnames]{xcolor}
\usepackage{color}
\usepackage{multirow}
\usepackage{comment}
\usepackage{amssymb}
\usepackage[version=3]{mhchem}
\usepackage{orcidlink}

\DeclareUnicodeCharacter{2212}{-}

\allowdisplaybreaks

\bibpunct{[}{]}{,}{n}{}{,}
\definecolor{darkspringgreen}{rgb}{0.09, 0.45, 0.27}

\begin{document}

\title{Galaxy clusters as cosmic lenses for high frequency gravitational-waves}

\author{Francesca Lecce~\orcidlink{0009-0009-6967-5042}}
\email{francesca.lecce@ba.infn.it}
\affiliation{Dipartimento Interuniversitario di Fisica  ``Michelangelo Merlin'', Via Amendola 173, 70126 Bari, Italy}
\affiliation{Istituto Nazionale di Fisica Nucleare - Sezione di Bari, Via Orabona 4, 70126 Bari, Italy}
\affiliation{CAPA \& Departamento de F\'isica Te\'orica, Universidad de Zaragoza, C. Pedro Cerbuna 12, 50009 Zaragoza, Spain}

\author{Francisco R. Cand\'on~\orcidlink{0009-0002-3199-9278}}
\email{francandon@unizar.es}
\affiliation{Fakult\"at f\"ur Physik, TU Dortmund, Otto-Hahn-Str. 4, Dortmund D-44221, Germany}
\affiliation{CAPA \& Departamento de F\'isica Te\'orica, Universidad de Zaragoza, C. Pedro Cerbuna 12, 50009 Zaragoza, Spain}

\author{Alessandro Lella~\orcidlink{0000-0002-3266-3154}}
\email{alessandro.lella@unipd.it}
\affiliation{Dipartimento di Fisica e Astronomia, Università degli Studi di Padova, Via Marzolo 8, 35131 Padova, Italy}
\affiliation{Istituto Nazionale di Fisica Nucleare (INFN), Sezione di Padova, Via Marzolo 8, 35131 Padova, Italy}%

\author{Alessandro Mirizzi~\orcidlink{0000-0002-5382-3786}}
\email{alessandro.mirizzi@ba.infn.it}
\affiliation{Dipartimento Interuniversitario di Fisica  ``Michelangelo Merlin'', Via Amendola 173, 70126 Bari, Italy}
\affiliation{Istituto Nazionale di Fisica Nucleare - Sezione di Bari, Via Orabona 4, 70126 Bari, Italy}%

\author{Julia K. Vogel}
\email{julia.vogel@tu-dortmund.de}
\affiliation{Fakult\"at f\"ur Physik, TU Dortmund, Otto-Hahn-Str. 4, Dortmund D-44221, Germany}

\author{Tiziano Zanzarella~\orcidlink{0009-0006-7860-3882}}
\email{tiziano.zanzarella@unito.it}
\affiliation{Dipartimento di Fisica, Universit\`{a} di Torino, via P. Giuria 1, I--10125 Torino, Italy}
\affiliation{Istituto Nazionale di Fisica Nucleare, Sezione di Torino, via P. Giuria 1, I--10125 Torino, Italy}%

\date{\today}
\smallskip

\begin{abstract}
    High-frequency Gravitational Waves~(HFGW) ($f \gtrsim 1$ MHz) offer a distinctive probe of exotic physics. In the absence of dedicated detectors targeting this frequency range, indirect probes based on graviton–photon conversions in astrophysical magnetic fields become particularly relevant. In this work we show that the extended magnetic fields permeating Galaxy Clusters may act as cosmic lenses, leading to a strong enhancement of graviton-photon conversions. In this context, X- and $\gamma$-ray observations of the Virgo Cluster  through large field-of-view instruments such as eROSITA and COMPTEL yield novel limits on the amplitude of stochastic HFGW backgrounds in the frequency range $10^{17}-10^{22}$~Hz, improving upon constraints derived from the Galactic magnetic field by one to two orders of magnitude.
\end{abstract}

\maketitle

\section{Introduction}
\label{sec:intro}
High Frequency Gravitational Waves (HFGWs) are typically defined as signals with frequency exceeding the kHz frequency range. While standard astrophysical processes may approach this threshold, none are expected to generate a significant gravitational-wave background in MHz or GHz domains. Consequently, the detection of a signal at these frequencies is considered a smoking gun for exotic physics or early-Universe cosmological processes~\cite{Rosado:2011kv,Sesana:2016ljz,Lamberts:2019nyk,Robson:2018ifk,KAGRA:2021kbb,KAGRA:2021vkt,Babak:2023lro}. Potential sources of HFGWs are generally classified into two broad categories: early-Universe and late-Universe mechanisms~(see Ref.~\cite{Aggarwal:2025noe} for a recent review). The former includes processes such as inflation~\cite{Grishchuk:1974ny,Starobinsky:1979ty,Rubakov:1982df,Bartolo:2016ami}, radiation from the plasma in the early-Universe~\cite{Ringwald:2020ist,Ringwald:2022xif}, first-order cosmological phase transition~\cite{Caprini:2015zlo,Caprini:2019egz,Kibble:1981gv,Gouttenoire:2022gwi,Athron:2023xlk}, topological defects~\cite{Vilenkin:2000jqa,Blanco-Pillado:2017oxo,Wang:2022pav,Gouttenoire:2019kij,Servant:2023tua}, and the evaporation of Primordial Black Holes~(PBHs)~\cite{Anantua:2008am,Dolgov:2011cq,Dong:2015yjs,Gehrman:2022imk,Gehrman:2023esa, Ireland:2023avg, Ireland:2023zrd, Profumo:2026suy, Su:2026agc}. In contrast, late-Universe sources are associated with exotic objects or specific high-energy processes, such as the merger of boson stars and light PBHs~\cite{Visinelli:2021uve, Bezares:2022obu, Franciolini:2022htd, Ge:2024itl, Chang:2024xjp}, black hole super-radiance~\cite{Brito:2015oca}, or nuclear processes within stars~\cite{Garcia-Cely:2024ujr} (see Ref.~\cite{Aggarwal:2020olq} for a review). Probing gravitational waves above the MHz range is significantly beyond the sensitivity limits of current interferometers such as LIGO, Virgo and Kagra~\cite{LIGOScientific:2014qfs,LIGOScientific:2019vic}, and remains out of reach for future ground-based observatories like the Einstein Telescope~\cite{Hild:2010id,Punturo:2010zz} and Cosmic Explorer~\cite{LIGOScientific:2016wof}. At these high frequencies, the signal begins to exhibit a particle-like nature, where it is more accurately described in terms of its quanta, gravitons. Consequently, the physical principle underlying detection must be adapted, requiring the development of indirect detection techniques.
One of the most promising approaches exploits the inverse Gertsenshtein effect~\cite{Gertsenshtein:1962kfm,Macedo:1983wcr,Raffelt:1987im,Cruise:2012zz,Dolgov:2012be,Ejlli:2018hke,Ejlli:2019bqj,Ejlli:2020fpt,Cembranos:2023ere}, i.e., the conversion of GWs into photons within an external magnetic field. This phenomenon is physically analogous to the axion-photon conversion mechanism used to probe axions~\cite{Raffelt:1987im}. 

As a result, experimental {designs developed} for axion searches can be directly optimized for dual-use GW detection. Upper limits on stochastic GW backgrounds at optical and X-ray frequencies were established utilizing data from light-shining-through-a-wall  experiments such as ALPS and OSQAR, alongside the CAST axion helioscope~\cite{Ejlli:2019bqj}, with upcoming expansions planned via ALP II~\cite{Bahre:2013ywa} and (Baby)IAXO~\cite{Armengaud:2014gea}. {Beyond static laboratory magnetic fields, alternative setups exploit the inverse Gertsenshtein effect using high-energy pulsed lasers in the optical band~\cite{Vacalis:2023gdz}, or probe  GW-induced polarization within microwave waveguides~\cite{Cruise:2000za} and optical cavities~\cite{Cruise:2005uq,Barrabes:2010tr}. Furthermore, axion dielectric haloscopes acting around $10\,\rm{GHz}$, such as the MADMAX~\cite{MADMAX:2019pub} and the DALI~\cite{DeMiguel:2020rpn,DeMiguel:2023nmz} experiments, have also been adapted for HFGW searches.}

Complementing ground-based experiments, cosmic magnetic fields offer an intriguing astrophysical laboratory to observe conversion signatures. Examples of these environments include planetary and stellar magnetospheres~\cite{Liu:2023mll,Ito:2023nkq, Dandoy:2024oqg, McDonald:2024nxj}, the Milky-Way (MW) magnetic field~\cite{Ito:2023nkq,Ramazanov:2023nxz,Lella:2024dus}, primordial extragalactic fields~\cite{Domcke:2020yzq} and the large scale magnetic field within galaxy clusters~\cite{He:2023xoh}.
The strongest bounds on the diffuse HFGW background over a broad frequency range $\sim10^9$--$10^{16}\,\rm{Hz}$ have been derived by exploiting conversions in the Galactic magnetic field \cite{Lella:2024dus}. On the other hand, it has recently been realized that several astrophysical environments, either hosting strong magnetic fields (e.g., starburst galaxies) or extending over much larger scales than the MW (e.g., galaxy clusters) may substantially enhance axion-photon conversions~\cite{Ning:2024eky,Ning:2025tit,Ning:2025kyu,Lecce:2025vjc}, thereby probing unexplored regions of the parameter space. Extending the analogy between axion- and graviton-photon conversions, these environments may act as powerful \emph{cosmic lenses}, converting the HFGW background traversing them into an observable photon signal. In this regard, the Virgo Galaxy Cluster~(VC) offers an ideal environment for conducting such searches~\cite{Ning:2024eky,Lecce:2025vjc,Gupta:2026scx}. Although its magnetic field strength, $B\sim0.1\,\mu{\rm G}$, is weaker than that of the MW, the extent of the magnetic domain $L\sim{\mathcal O}(10^3)~{\rm kpc}$--100 times larger than the galactic one-- dramatically enhances the volume available to HFGW conversions. Large-field-of-view telescopes targeting different frequency ranges, such as eROSITA~\cite{eROSITA:2020emt} and COMPTEL~\cite{Strong:1998ck}, are therefore particularly well suited for these searches. Indeed, the larger number of signal photons collected over the extended cluster region observed is accompanied by only a modest increase in the photon background, since most stellar emission is concentrated toward the cluster center.

The paper is structured as follows. In Sec~\ref{sec:Graviton-photonconversion} the theoretical framework associated with the Gertsenshtein effect is reviewed. Sec.~\ref{sec:electrondensityandmagneticfield} discusses the modeling of the VC environment. Sec.~\ref{sec:conversionsinVirgo} presents the typical conversion probabilities derived from the cluster model. Sec.~\ref{sec:experiment} outlines the multi-wavelength observations of the VC, encompassing the X-ray regime covered by eROSITA~\cite{eROSITA:2020emt} and NuSTAR~\cite{NuSTAR:2013yza}, as well as the soft $\gamma$-ray band observed by COMPTEL~\cite{Strong:1998ck}. In Sec.~\ref{sec:bounds}, the exclusion limits obtained from these observations are presented. Finally, in Sec.~\ref{sec:conclu} we summarize our results and conclude.

\section{Graviton-photon conversion}
\label{sec:Graviton-photonconversion}
The conversion of gravitons into photons in the presence of an external magnetic field, known as the \textit{inverse Gertsenshtein effect}, can be characterized through the following action~\cite{Dolgov:2012be,Ejlli:2018hke}
\begin{equation}
    S=\int{d^4x\sqrt{-g}\,\left(\mathcal{L}_{\rm gr}+\mathcal{L}_{\rm em}\right)}\, .
    \label{eq:action}
\end{equation}
This formulation adopts natural units  $(\hbar=c=1)$ and assumes a Minkowski background with a metric signature $\eta=\rm{diag}(1,\,-1,\,-1,\,-1)$. Here, $\sqrt{-g}$ denotes the determinant of the curved spacetime metric $g_{\mu\nu}(x)$. To study weak gravitational interactions, the metric is expanded as
\begin{equation}
    g_{\mu\nu}(x)=\eta_{\mu\nu}+\frac{\sqrt{2}}{M_P}h_{\mu\nu}(x)\,,
\end{equation}
where $M_P=\left(8\pi G\right)^{-1/2}\approx 2.44\times 10^{18}\,\rm{GeV}$ is the reduced Planck mass and $h_{\mu\nu}(x)$ is a linear perturbation propagating on the flat space time metric $\eta_{\mu\nu}$, under the assumption that the fluctuations $h_{\mu\nu}$ are extremely small. The first term in the action describes the free-graviton Lagrangian, given by the Einstein-Hilbert term
\begin{equation}
    \mathcal{L}_{\rm gr}=\frac{M_{P}^2}{2}R\,,
\end{equation}
where $R$ denotes the Ricci scalar curvature. \\
The second term, $\mathcal{L}_{\rm em}$, describes the dynamics of the electromagnetic field minimally coupled to the gravitational background
\begin{equation}
    \begin{split}
        \mathcal{L}_{\rm em}=&-\frac{1}{4}g^{\mu\alpha}g^{\nu\beta}F_{\mu\nu}F_{\alpha\beta}\\
        &-\frac{1}{2}\int d^4x' A_{\mu}(x)\Pi^{\mu\nu}(x,x')A_\nu(x')\,.
    \end{split}
\end{equation}
In this expression, $F_{\mu\nu}=\partial_\mu A_\nu-\partial_\nu A_\mu$ is the electromagnetic field strength and $\Pi^{\mu\nu}(x,x')$ the photon polarization tensor describing propagation in a magnetized medium.
For a beam propagating along the $z$-axis (wave vector $\mathbf{k}\parallel\hat{\mathbf{z}}$) with energy $\omega$, the fields can be decomposed into their respective polarization states.
In the Coulomb gauge, the vector potential is written as
\begin{equation}
    A_{i}(\mathbf{x},t)=\sum_{\lambda=x,y}A_{\lambda}(\mathbf{x})\epsilon_{i}^\lambda e^{-i\omega t}\,,
\end{equation}
where $\epsilon_{i}^\lambda$ are the linear polarization unit vectors in the plane transverse to the direction of motion $z$.  Analogously, in the Transverse-Traceless (TT) gauge, the gravitational perturbation takes the form
\begin{equation}
    h_{ij}(\mathbf{x},t)=\sum_{\sigma=\times,+}h_{\sigma}(\mathbf{x)}e_{{ij}}^\sigma\,e^{-i\omega t}\,.
\end{equation}
The polarization tensors $\{e_{{ij}}^\times,e_{ij}^+\}$ are constructed such that 
\begin{equation}
\begin{split}
e_{{ij}}^\times&=\epsilon_{i}^x\epsilon_{j}^y+\epsilon_{i}^y\epsilon_{j}^x \,,\\   e_{ij}^+&=\epsilon_{i}^x\epsilon_{j}^x-\epsilon_{i}^y\epsilon_{j}^y\,.
\end{split}
\end{equation}
Applying the Euler-Lagrange equations to the action shown in Eq.~\eqref{eq:action} yields the coupled equations of motion (EoM) for the system. 
{For a graviton--photon beam propagating along the $z$-axis, one obtains~\cite{Gertsenshtein:1962kfm,Raffelt:1987im,Ejlli:2018hke}}

\begin{equation}
\left(i\frac{d}{dz}-\omega\right)
    \begin{pmatrix}
        h_+\\ h_\times \\ A_{x} \\ A_{y}
    \end{pmatrix}=H\begin{pmatrix}
        h_+\\ h_\times \\ A_{x} \\ A_{y}
    \end{pmatrix}\,.
    \label{eq:eom}
\end{equation}
{This Schrödinger-like form is convenient for a simple one dimensional numerical integration along the Line of Sight~(LoS)}. As established in Refs.~\cite{Millar:2021gzs,Satherley:2024wsz,McDonald:2024uuh}, a rigorous three-dimensional treatment would reveal the possibility of resonant conversions when the beam traverses resonance surfaces, thereby enhancing the associated conversion probability. For the sake of simplicity, we do not include these effects in our treatment, being aware that they would only strengthen our results.

The effective Hamiltonian $H$ can be written in forms of blocks encoding the particle dispersion relation and the mixing effect between the fields
\begin{equation}
  H=\begin{pmatrix}
        0 & \mathcal{H}_{g\gamma}\\
        \mathcal{H}_{g\gamma} & \mathcal{H}_{\gamma\gamma}
    \end{pmatrix}\,.
    \label{eq:Hmatrix}
\end{equation}
The matrix $2\times2$ $\mathcal{H}_{\gamma\gamma}$ encodes the photon dispersion relation. In the energy range considered in this work, the photon birefringence term and the CMB terms are negligible, and the photon dispersion matrix can be written as~\cite{Lella:2024dus}
\begin{equation}
    \mathcal{H}_{\gamma\gamma}=\begin{pmatrix}
        \Delta_{\rm{pl}} & 0\\
         0 & \Delta_{\rm{pl}}
    \end{pmatrix}\,,
\end{equation}

with 
\begin{equation}
\begin{split}
    \Delta_{\rm{pl}}&=-\frac{\omega_{\rm{pl}}^2}{2\omega}\\
    &\simeq -2.7 \times 10^{15}\left(\frac{1\,\rm{Hz}}{f}\right)\left(\frac{n_e}{10^{-4}\,\rm{cm}^{-3}}\right)\,\rm{kpc}^{-1}\,,
\end{split}
\label{eq:deltapl}
\end{equation}
where $\omega_{\rm pl}= \sqrt{4 \pi\alpha n_e/m_e}$ is the plasma frequency, $n_e$ and $m_e$  are the electron number density and electron mass, respectively. Finally, $f$ is the GW frequency.
 {The off-diagonal term in Eq.~(\ref{eq:Hmatrix}) term accounts for the graviton-photon interaction and takes the form}

\begin{equation}
    \mathcal{H}_{g\gamma}=\begin{pmatrix}
        \Delta_{g\gamma}\sin\phi &   \Delta_{g\gamma}\cos\phi\\
     \Delta_{g\gamma}\cos\phi& -  \Delta_{g\gamma}\sin\phi
    \end{pmatrix}\,,
\end{equation}
where
\begin{equation}
    \Delta_{g\gamma}=\frac{B_T}{\sqrt{2}M_P}\simeq 8.8\times 10^{-11}\left(\frac{B_T}{10^{-7}\, \rm G}\right)\,\rm{kpc}^{-1}
     \label{eq:deltaggamma}\,
\end{equation}
{and $\phi$ is the angle between the transverse component of the magnetic field $B_T$ and the orientation of the polarization vector $\epsilon_x$.
}
For an initially unpolarized graviton beam, the conversion probability after a distance $z$ is obtained by taking the average over the two independent initial graviton polarization states ($h_{+}$ and $h_{\times}$)
\begin{equation}
    P_{g\gamma}=\frac{1}{2}\left(|A_{x}(z)|^2+|A_y(z)|^2\right)\,.
\end{equation}
Since the Hamiltonian generally does not commute at different points along the LoS, the system does not admit a general analytical solution. In our analysis, we employ a perturbative approach to solve the equations of motion, {which matches the results obtained by solving the full evolution equations for magnetic environments defined on a finite-resolution spatial grid; we outline the details of this treatment and discuss its validity in Appendix~\ref{app:perturbative}. We also investigate the effect of hard edges of the profiles used to compute the conversion probability. We discuss the robustness of our results against the smoothing of the magnetic field cell edges in Appendix~\ref{app:hardedges}.}

\section{Graviton-photon conversions in Virgo Cluster magnetic fields}

\subsection{Plasma and magnetic fields in the Virgo Cluster}
\label{sec:electrondensityandmagneticfield}
To model the magnetic field and the electron density of the VC, simulation data from \textsc{the IllustrisTNG} project were used. The set of cosmological magneto-hydrodynamical models of galaxy formation provides predictions for both current and future observational programs~\cite{Nelson:2019jkf,Marinacci:2017wew}. 
{We adopt the TNG300-1 realization as the benchmark model, since it features the highest spatial resolution available for the largest possible simulation volume. Within the simulation volume} {we identify subhalos compatible with the total mass of the VC, $M = (6.3 \pm 0.9) \times 10^{14} M_\odot$~\cite{Ning:2024eky}}. Seven candidates have been found and modeled as spheres with a radius of $1\,\rm{Mpc}$. In order to assess the uncertainty, we perform the full analysis with all the compatible magnetic field models.

\begin{figure}[t]
    \centering
    \includegraphics[width=1\columnwidth]{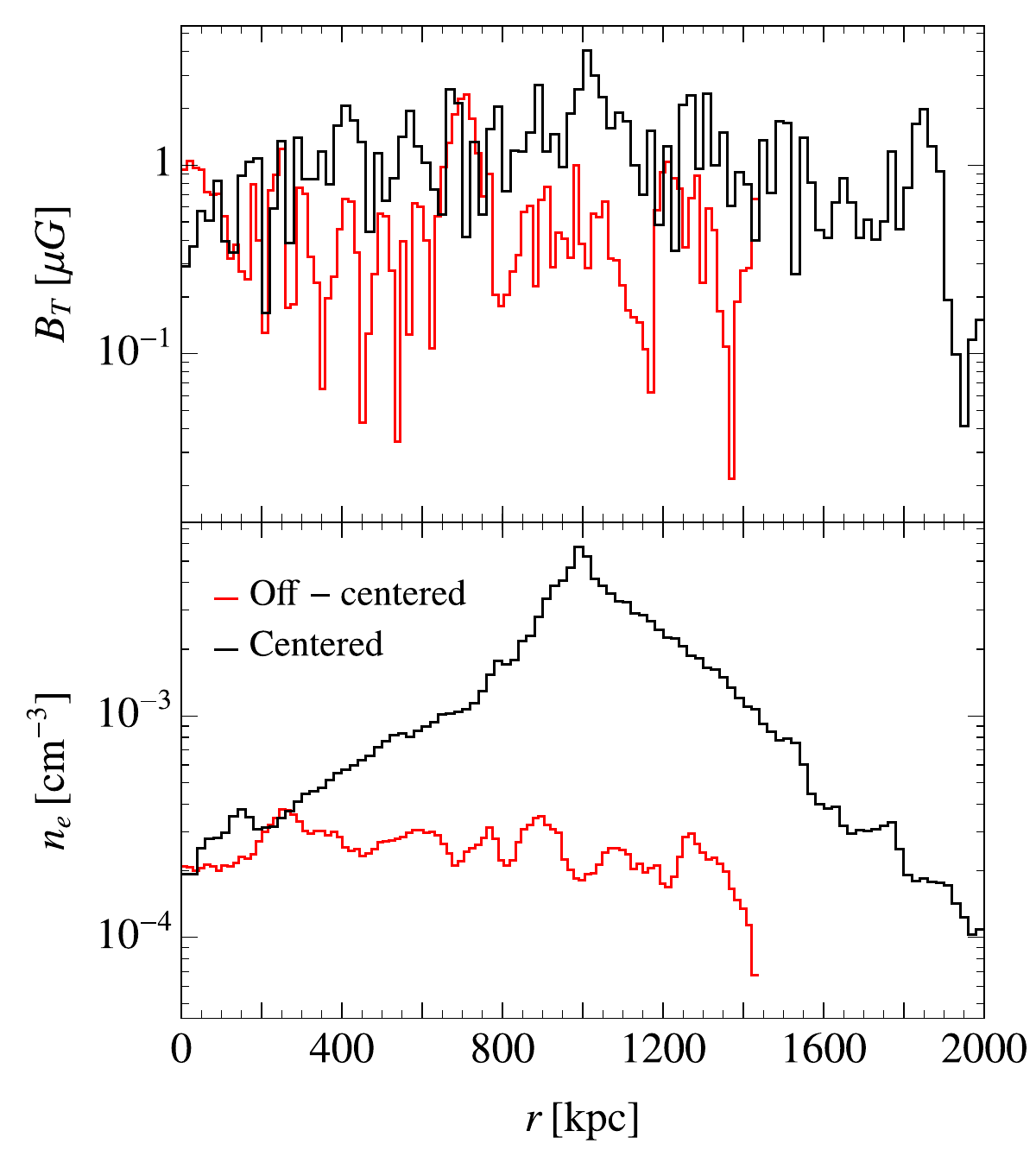}
    \caption{Profiles of the transverse magnetic field $B_T$ (upper panel) and the free electron density $n_e$ (lower panel) as a function of the distance traveled within the cluster $r$ for a representative centered~(black line) and off-centered~(red line) trajectory. A 3D visualization of the chosen LoSs is provided in Fig.~\ref{fig:sphereplot}. }
    \label{fig:plotneBTcomb}
\end{figure}

\begin{figure}
    \centering
    \includegraphics[width=0.8\columnwidth]{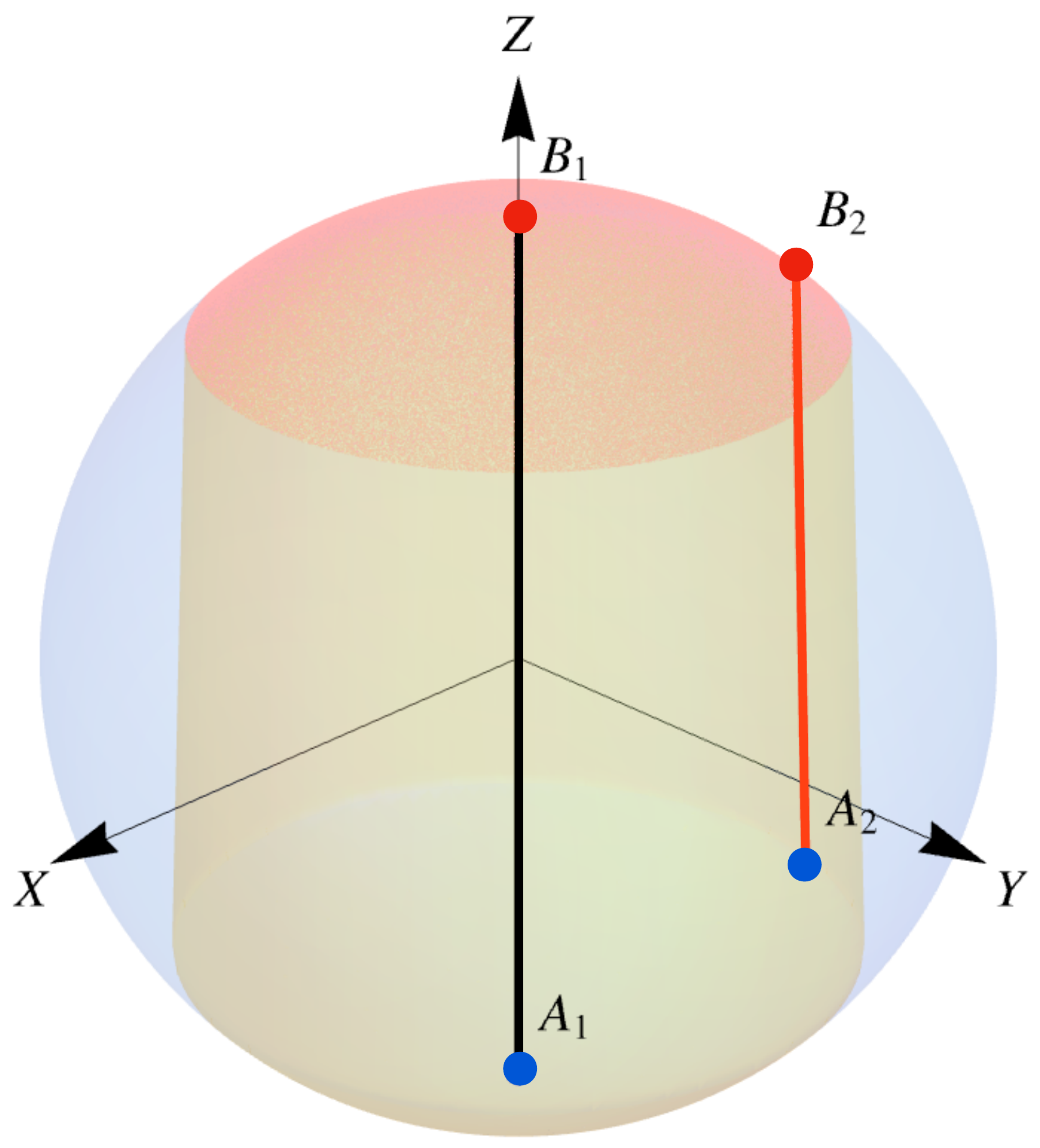}
    \caption{
    3D visualization of the geometric configuration of the two representative profiles shown in Fig.~\ref{fig:plotneBTcomb}. The yellow region represents the volume subtended by the $2.5^\circ$ FoV of a generic instrument, while the red cap corresponds to the observed surface. The black line indicates a trajectory passing through the cluster center and aligned with $z$-axis of our reference frame. It enters the cluster volume in the point $A_1$ and exits from the point $B_1$. The off-centered trajectory is represented by the red line connecting the entry point $A_2$ to the exit point $B_2$.}
     \label{fig:sphereplot}
\end{figure}

HFGWs traveling through the cluster volume towards the detector may follow either radial trajectories passing through the simulation frame origin or off-centered trajectories. Depending on the trajectory, HFGW may traverse substantially different magnetic-field and electron-density profiles, with regions closer to the cluster center characterized by stronger fields and higher densities. In Fig.~\ref{fig:plotneBTcomb} we display the magnetic field (top panel) and electron density (bottom panel) profiles for representative 
centered and off-centered trajectories. A 3D visualization of the selected LoSs is provided in Fig.~\ref{fig:sphereplot}, where the host environment is modeled as  sphere of radius $R$. The yellow region displays the volume of the environment observed by a representative instrument featuring a $2.5^\circ$ opening Field of View (FoV),  while the red cap represents the emitting surface visible to the experiment. The centered trajectory, shown as a black line along the $z$-axis in Fig.~\ref{fig:sphereplot}, originates in the hemisphere opposite the observer surface at $A_1$ and spans the entire diameter of the sphere, exiting the cluster volume at $B_1$. For an off-centered trajectory within the observer FoV (shown as a red line), the HFGW flux enters the environment at an off-axis point $A_2$, and exits the domain at the point $B_2$. In Appendix~\ref{app:geometry}, we illustrate in detail the geometric parametrization chosen to describe the different propagation directions.

It is worth noting that alternative models of the VC's magnetic field have been proposed in the literature. A comparison between the results obtained using \textsc{IllustrisTNG} and those from the model presented in Ref.~\cite{Marsh:2017yvc} is provided in Appendix~\ref{app:Marsh}.

\subsection{Photon flux from HFGW conversions}
\label{sec:conversionsinVirgo}
\begin{figure}
    \centering
    \includegraphics[width=1\columnwidth]{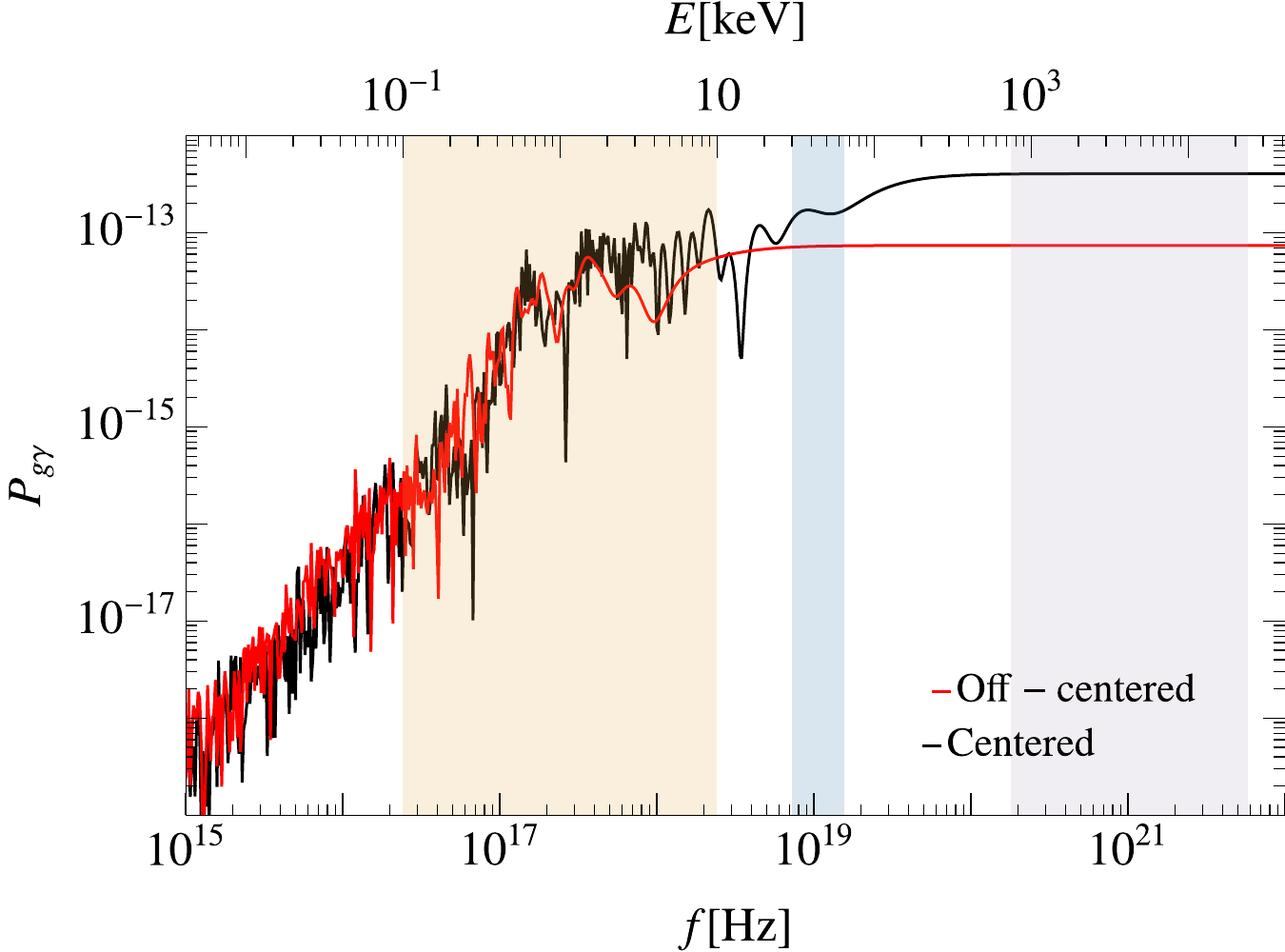}
    \caption{Graviton-to-photon conversion probability $P_{g\gamma}$ as a function of frequency $f$~(lower axis) and energy $E$~(upper axis) for the two representative centered~(black line) and off-centered~(red line) LoSs traversing the VC. These probabilities are calculated using the magnetic field and electron density profiles shown in Fig.~\ref{fig:plotneBTcomb}. The orange shaded band indicates the frequency range in which eROSITA is operational, the blue shaded region denotes the frequency range for NuSTAR data as reported in Ref.~\cite{Ning:2024eky}, while the violet shaded one refers to the range of COMPTEL data as reported in Ref.~\cite{Manna:2024tgu}.}
    \label{fig:plotPgg}
\end{figure}

The graviton-to-photon conversion probability $P_{g\gamma}$ is evaluated by solving the equations of motion Eq.~\eqref{eq:eom} along the LoSs traversing the VC volume. To directly relate the conversion efficiency to the astrophysical profiles discussed in Sec.~\ref{sec:electrondensityandmagneticfield}, the probability is evaluated for the same trajectories: the centered path passing through the cluster center and the off-centered trajectory introduced before. The resulting frequency dependent conversion probabilities are shown in Fig.~\ref{fig:plotPgg}, where the black and red curves correspond to the profiles previously presented in Fig.~\ref{fig:plotneBTcomb}. These results illustrate the sensitivity of the conversion process to the underlying spatial variations in the magnetic field strength and electron density.

Furthermore, the shaded bands in Fig.~\ref{fig:plotPgg} indicate the experimental frequency range of interest for this work. 
In particular, the orange shaded band indicates the frequency range in which eROSITA is operational, the blue shaded band denotes the frequency range for NuSTAR data as reported in Ref.~\cite{Ning:2024eky}, while the violet shaded band refers to the range of COMPTEL data as reported in Ref.~\cite{Manna:2024tgu}.
In this setup, the VC is modeled as a sphere of radius $R=1\,\rm{Mpc}$ centered at the origin of our reference frame, with the observer positioned along the $z$-axis at a distance $D_{\rm VC}=16.3\,\rm{Mpc}$. Due to the geometric constraints of the FoV, only a spherical cap facing the detector contributes to the converted flux.
As derived in Appendix~\ref{app:geometry}, each exit point on the visible surface defines a unique LoS corresponding to a specific trajectory within the cluster.
An isotropic gravitational wave background is associated with an energy flux per unit frequency and solid angle, given by~\cite{Maggiore:2000gv,Dandoy:2024oqg}
\begin{equation}
    \frac{d^2F_g}{df d\Omega}=\frac{\pi}{2}M_{P}^2\,f\,h_c^2=2.4 \times 10^{64} \left(\frac{f}{\textrm{Hz}} \right) h_c^2  \,\ \textrm{cm}^{-2}\textrm{s}^{-1}\,\textrm{sr}^{-1}\,
\label{eq:gravitonflux}
\end{equation}
where $h_c$ is the characteristic strain defined as in Ref.~\cite{Maggiore:2000gv}, which parametrizes the strength of the GW signal. The induced photon energy flux per unit frequency and solid angle is then calculated by weighting the incident graviton flux by the conversion probability $P_{g\gamma}(f,\Omega)$
\begin{equation}
    \frac{d^2F_{\gamma}}{df d\Omega}(f,\Omega)=\frac{d^2F_{g}}{df d\Omega}P_{g\gamma}(f,\Omega)\,\ .
\end{equation}
Ultimately, the total flux is computed numerically by integrating the contributions from the sampled lines of sight across the whole spherical cap defined by the instrument FoV. We refer the reader to Appendix~\ref{app:geometry} for further details.

\section{Observations of the Virgo Cluster}
\label{sec:experiment}

The VC is the closest large galaxy cluster and its center coincides with the position of Messier 87 (M87), a giant elliptical galaxy hosting a supermassive black hole at coordinates $\text{RA} = 187^\circ\,42'\,21.3'', \, \text{Dec} = +12^\circ\,23'\,28.0''$.
Viewed from Earth, the cluster subtends a significant portion of the sky, with an angular radius of approximately $4^\circ$ extending from the galaxy M87.
Traditionally, galaxy clusters have been observed across a broad range of the electromagnetic spectrum, from radio wavelengths to hard X-rays. 
In this work, the VC is analyzed using data from three distinct observatories covering different energy ranges: eROSITA, NuSTAR and COMPTEL~\cite{Collmar:1999rc}. The primary technical characteristic of these experiments are shown in Tab.~\ref{tab:Tab1}.
A particularly crucial parameter in this analysis is the FoV of the experiments. Since the angular extent of the VC ($4^\circ$) exceeds the FoV of several instruments, the visible volume, and consequently the integrated signal, is limited by the instruments' aperture. The specific data utilized and the regional processing for each instrument are detailed in the following Subsections.

\begin{table*}[t!]
\caption{
Technical parameters of the instruments considered in this study, including the energy range, characteristic frequency range, typical effective area, and FoV. The quoted FoV corresponds to the instantaneous FoV and should be distinguished from the survey sky coverage. In survey mode, eROSITA and COMPTEL provide all-sky coverage.
}
\centering
\begin{tabular}{c c c c c}
\hline
Experiment &
Energy range &
Frequency range &
Effective area &
FoV radius\\
&
$[\mathrm{keV}]$ &
$[\mathrm{Hz}]$ &
$[\mathrm{cm}^{2}]$ &
 $[^\circ]$
\\
\hline
\hline

eROSITA &
$0.2$--$10$ &
$10^{16}$--$10^{18}$ &
$\sim 1500$ at $1.5~\mathrm{keV}$ &
$\sim 0.52^\circ$ \\

NuSTAR &
$30$--$70$\textsuperscript{\dag} &
$\sim 10^{19}$ &
$\sim 847$ at $9~\mathrm{keV}$ &
$ \sim 0.12^\circ$ \\

COMPTEL~\cite{Collmar:1999rc} &
$7\times 10^{2}$--$3\times 10^{4}$ &
$10^{20}$--$10^{22}$ &
$\sim 40$ at $5~\mathrm{MeV}$ &
$\sim 33^\circ$ \\

\hline
\end{tabular}
\begin{flushleft}
\small
\textsuperscript{\dag}~The energy range shown correspond to the one adopted in Ref.~\cite{Ning:2024eky}. The nominal energy range of NuSTAR is $3$--$79~\mathrm{keV}$.
\end{flushleft}
\label{tab:Tab1}
\end{table*}

\subsection{eROSITA}

The extended Roentgen Survey and Imaging Telescope Array (eROSITA) \cite{eROSITA:2020emt} was launched aboard the Spektrum Roentgen Gamma (SRG) mission \cite{eROSITA:2021bln} in 2019. We analyzed and processed public available eROSITA observations of the VC from the first data release (DR1). eROSITA is particularly suited to this purpose because of its large FoV and sensitivity to faint, spatially extended X-ray emission in the soft X-ray band (See Tab. \ref{tab:Tab1}). These properties have already enabled eROSITA wide-field studies of the VC and its outskirts \cite{eROSITA:2024pin}.

\begin{figure}[h]
    \centering
    \includegraphics[width=1\columnwidth]{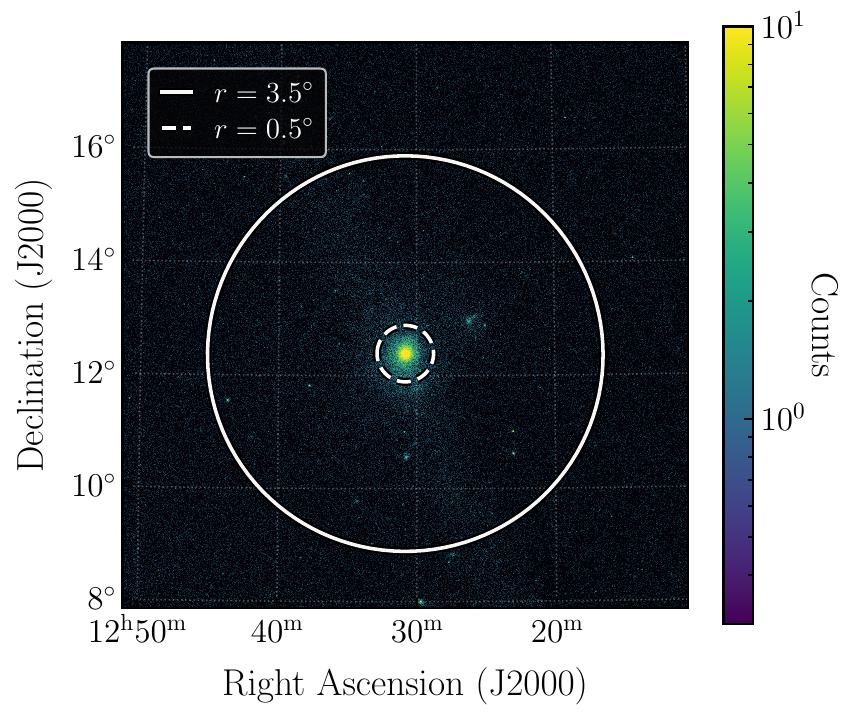}
    \caption{eROSITA count map of the VC centered on M87 for the 0.2-10.0 keV energy range. The solid and dashed circles show the extraction regions of radius $r = 3.5^\circ$ and $r = 0.5^\circ$  used in this work. The counts scale maximum is fixed to 10 counts for visualization purposes. }
    \label{fig:FoV}
\end{figure}

The VC skytiles used in this analysis were retrieved from the eROSITA-DE Data Release 1 archive \cite{eROSITADE:DR1eRODat} using M87 coordinates $\text{RA} = 187^\circ\,42'\,21.3'', \, \text{Dec} = +12^\circ\,23'\,28.0''$ and a maximum radius of $10^\circ$. The event processing was performed with the eROSITA Science Analysis Software System, \texttt{eSASS}. The event files were first merged using \texttt{evtool}. Following the same analysis techniques applied in \cite{eROSITA:2024pin}, we selected single, double, triple, and quadruple events patterns by imposing \texttt{pattern=15}, and applied the standard flag selection \texttt{flag=0xe00fff30} to reject bad pixels and strongly vignetted detector regions. Periods affected by enhanced particle background were removed with \texttt{flaregti}, using high-energy events above $5~{\rm keV}$ to identify flaring intervals. The resulting good-time intervals were then applied to produce the cleaned event file. The data were centered on the M87 coordinates using \texttt{radec2xy}, and the final cleaned event file was rebinned with a pixel size of $6.2''$. For visualization, we produced an eROSITA image in the $0.2$--$10~{\rm keV}$ energy range, together with the two circular source regions used in the analysis, $3.5^\circ$ and $0.5^\circ$, in Fig.~\ref{fig:FoV}. 

\begin{figure}[h]
    \centering
    \includegraphics[width=1\columnwidth]{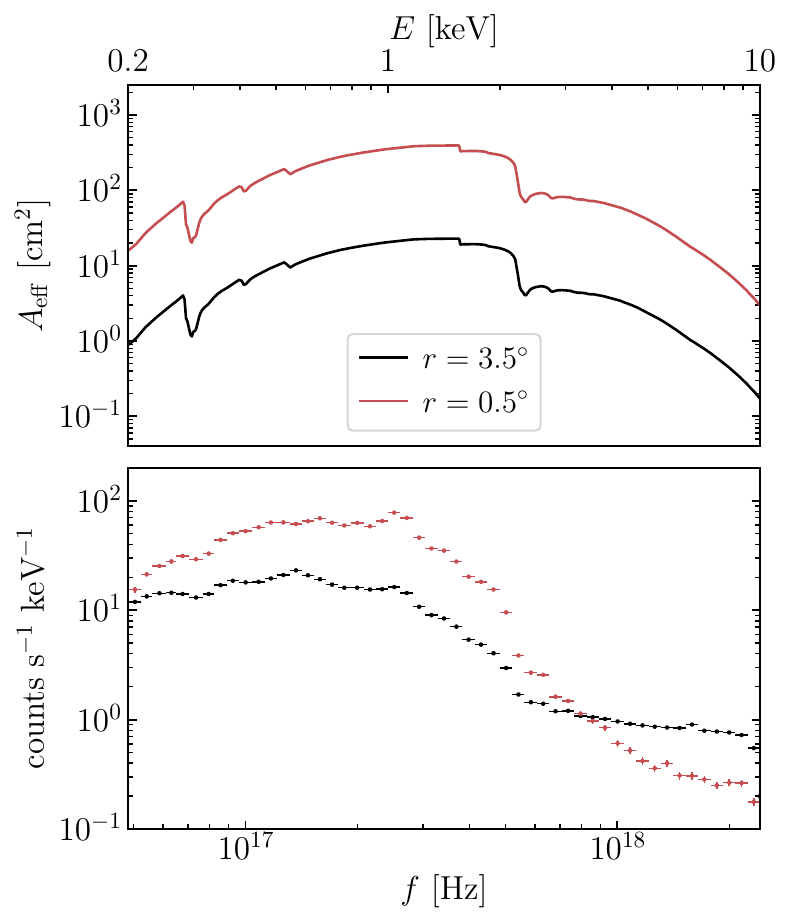}
    \caption{Upper panel: Effective area obtained from the eROSITA response files for the $0.5^\circ$ and $3.5^\circ$ circular extraction regions. Lower panel: Observed eROSITA spectra of the VC for the $3.5^\circ$ and $0.5^\circ$ circular extraction regions. No astrophysical or instrumental background subtraction is applied.}
    \label{fig:spectra}
\end{figure}

Spectra and instrumental response files were extracted with \texttt{srctool}. For each of the two circular apertures, \texttt{srctool} was used to generate the source spectrum, ancillary response file (ARF), and redistribution matrix file (RMF). Of the seven eROSITA telescope modules, TM5 and TM7 were excluded from the analysis because of optical light-leak contamination, which primarily affects the lower energy range from 0.2 to 0.8 keV~\cite{eROSITA:2024pin}. Therefore, this analysis uses only the combined TM8 data product, comprising TM1, TM2, TM3, TM4, and TM6. The effective area $A_{\mathrm{eff}}$ and the observed count-rate spectra for the two source regions are shown as a function of energy in Fig.~\ref{fig:spectra}. The effective area inferred for the $3.5^\circ$ extraction region is about one order of magnitude lower than that obtained for the $0.5^\circ$ region. This is mainly a geometrical and instrumental effect associated with the large angular size of the selected aperture. The $3.5^\circ$ region is much larger than the instantaneous eROSITA FoV, which is of order $\sim 1^\circ$. Therefore, during an individual survey scan, only a fraction of the selected aperture is observed at a given time, while large parts of the region have zero instantaneous exposure. In the final survey product, the exposure is accumulated over many scan passages, but it remains highly non-uniform across such a large region. Moreover, different parts of the aperture are observed at different off-axis angles and with different vignetting, detector coverage, and effective exposure. These effects are included in the response extraction and lead to a lower exposure-weighted effective area for the larger extraction region.
The corresponding spectra, in units of counts~s$^{-1}$~keV$^{-1}$, are shown in the lower panel of Fig.~\ref{fig:spectra}.
To remain conservative, we omit both astrophysical or instrumental background subtraction. Subtracting these components would reduce the residual count rate attributable to a potential signal; by leaving them in, we allow a larger fraction of the observed counts to be potentially ascribed to exotic emission. Accounting for these known background components would yield more stringent constraints, as explicitly demonstrated in Sec.~D of the Supplemental Material of Ref.~\cite{Candon:2024eah}.
Notably, the $3.5^\circ$ aperture lower count-rate spectrum is not simply an intrinsic property of the VC emission, but is also linked to the survey observing geometry that we mentioned above. In some passages the bright central region around M87 is not inside the FoV, while lower emission outer regions are observed. When the events are accumulated over the full survey extraction, the resulting spectrum is therefore weighted by a highly non-uniform exposure pattern. These effects are reflected in the much lower effective area obtained for the $3.5^\circ$ region.

\subsection{NuSTAR}

The \textsc{NuSTAR} Observatory~\cite{NuSTAR:2013yza} is the first focusing high-energy X-ray telescope launched into space. It consists of two co-aligned telescope modules, each equipped with independent optics and a CdZnTe focal-plane module, denoted FPMA and FPMB. The instantaneous FoV of each module is approximately $13' \times 13'$ ($0.22^\circ \times 0.22^\circ$), equivalent to a circular radius of $r_{\rm eq} \simeq 0.12^\circ$, and the on-axis angular resolution is $58''$ half-power diameter (HPD). As a result, a single \textsc{NuSTAR} observation covers only a small fraction of the VC, roughly limited to the central region around M87. In addition, since \textsc{NuSTAR} does not operate in survey mode, most observational campaigns target the same central region around M87, leaving the outskirts of the VC largely unobserved.

In this work, we use the archival \textsc{NuSTAR} data products presented in Ref.~\cite{Ning:2024eky}. In particular, we adopt their reduction and analysis of the M87 observations, corresponding to a total exposure of approximately $0.51~\mathrm{Ms}$ for each of the two focal-plane modules, FPMA and FPMB. From the extraction regions shown in Ref.~\cite{Ning:2024eky}, the source region used for M87 corresponds approximately to a circular aperture of radius $r_{\rm ON}\simeq 0.014^\circ$, or about $50''$, centered on M87.

\subsection{COMPTEL}
While previous sections focus on the X-ray emission of the VC, here we consider observations in the soft $\gamma$-ray band ($0.7-30$\,\rm{MeV}), using observations performed by Imaging Compton Telescope (COMPTEL) ~\cite{Collmar:1999rc}. COMPTEL was sensitive to $\gamma$-rays between $0.75$ and $30\,$MeV, featuring an energy resolution of $5$-$8\%$ and an angular resolution ranging from $1.7^\circ$ to $4.4^\circ$, depending on the incident photon energy. During its $9.7$ years of operation, the mission conducted approximately $340$ pointings, each of which typically lasted $2$ weeks and covered a wide FoV with a radius of approximately $30^\circ$.
The analysis presented here uses data extracted from~\cite{Manna:2024tgu}, in which pertinent viewing periods of the VC were processed within a Region of Interest (ROI) with an angular radius of $10^\circ$. Due to the relatively broad Point Spread Function (PSF) of the instrument, three distinct spatial templates were employed to characterize the source: a point source model, a radial disk, and a radial Gaussian. The point source model assumes emission originates from an infinitesimal point; the radial disk assumes a finite size with uniform brightness; and the radial Gaussian model assumes a finite size with an intensity profile that decreases from the center following a Gaussian distribution.
The Spectral Energy Distribution (SED) provided for these three models was used to derive the observed counts. Specifically, the reported flux values were multiplied by the total exposure reported as $1.2\times 10^{9}\,\rm{cm}^2\,\rm{s}$, and integrated over the respective energy bins to determine the number of photons per bin. This procedure provides the observed rate necessary for the derivation of the exclusion limits in the MeV range.
Regarding the spatial templates, the constraints shown by the violet region in Fig.~\ref{fig:bound} correspond to the point-source model. The bounds derived from the three alternative templates differ only by minor factors (roughly within $10-20\%$), demonstrating that our conclusions are robust against the choice of the template.
Notably, because the ROI used in this dataset is large enough to encompass the $1\,\rm{Mpc}$ sphere used in the magnetic field model, the resulting bounds are to be considered conservative. Indeed the theoretical graviton-to-photon emission from the cluster is compared against a dataset integrated over a much larger region of the sky, which includes not only the VC but also the surrounding spatial environment.
\section{Constraints on diffuse HFGW backgrounds}
\label{sec:bounds}
As established in Sec.~\ref{sec:conversionsinVirgo}, graviton-to-photon conversion results in an additional photon flux from the VC. To conservatively account for the intrinsic cluster emission, the observed flux is assumed to be entirely explained by non-exotic physics. Therefore, we require the flux arising from HFGW conversions to be fainter than the observed photon emission. This constraint directly translates into a limit on the abundance of the GW background, parameterized through the characteristic strain $h_c$.
For each of the detectors considered in Sec.~\ref{sec:experiment} graviton-to-photon conversion yields an expected number of detected photons per unit time

\begin{equation}
   R_{\Delta\!f}^{\rm th}(f)= \int_{f-\Delta f/2}^{f+\Delta f/2} df'\, \frac{1}{2 \pi f'} A_{\rm eff}(f')\int d\Omega \frac{d^2 F_\gamma}{df'\, d\Omega}(f',\Omega)
\end{equation}
within a frequency bin of width $\Delta f$ centered at frequency $f$.Specifically, the frequency bins corresponding to the energy bins used in our analysis for each instruments are structured as follows
\begin{itemize}
    \item eROSITA: $50$ logarithmically spaced energy bins spanning the $0.2–10\,\rm{keV}$ range.
    \item COMPTEL: $16$ logarithmically spaced energy bins spanning the $0.75-30$ MeV range.
    \item NuSTAR: $5$ keV wide energy bins spanning the $30-70$ keV range.
\end{itemize}

In this expression, $A_{\rm eff}(f)$ is the effective area of the instrument, and $\Omega$ is the solid angle subtended by the source within the aperture used to extract the signal. As described in Appendix~\ref{app:geometry}, if the angular extent of the source in the sky exceeds the telescope extraction aperture, $\Omega$ coincides with the solid angle of the extraction aperture itself (which is not, in general, the instantaneous detector FoV quoted in Table~\ref{tab:Tab1}). The expected number of signal photons is then compared to the observed event rate $R^{\rm exp}_{\Delta\!f}(f)$.

\begin{figure*}
    \centering
    \includegraphics[width=0.9\linewidth]{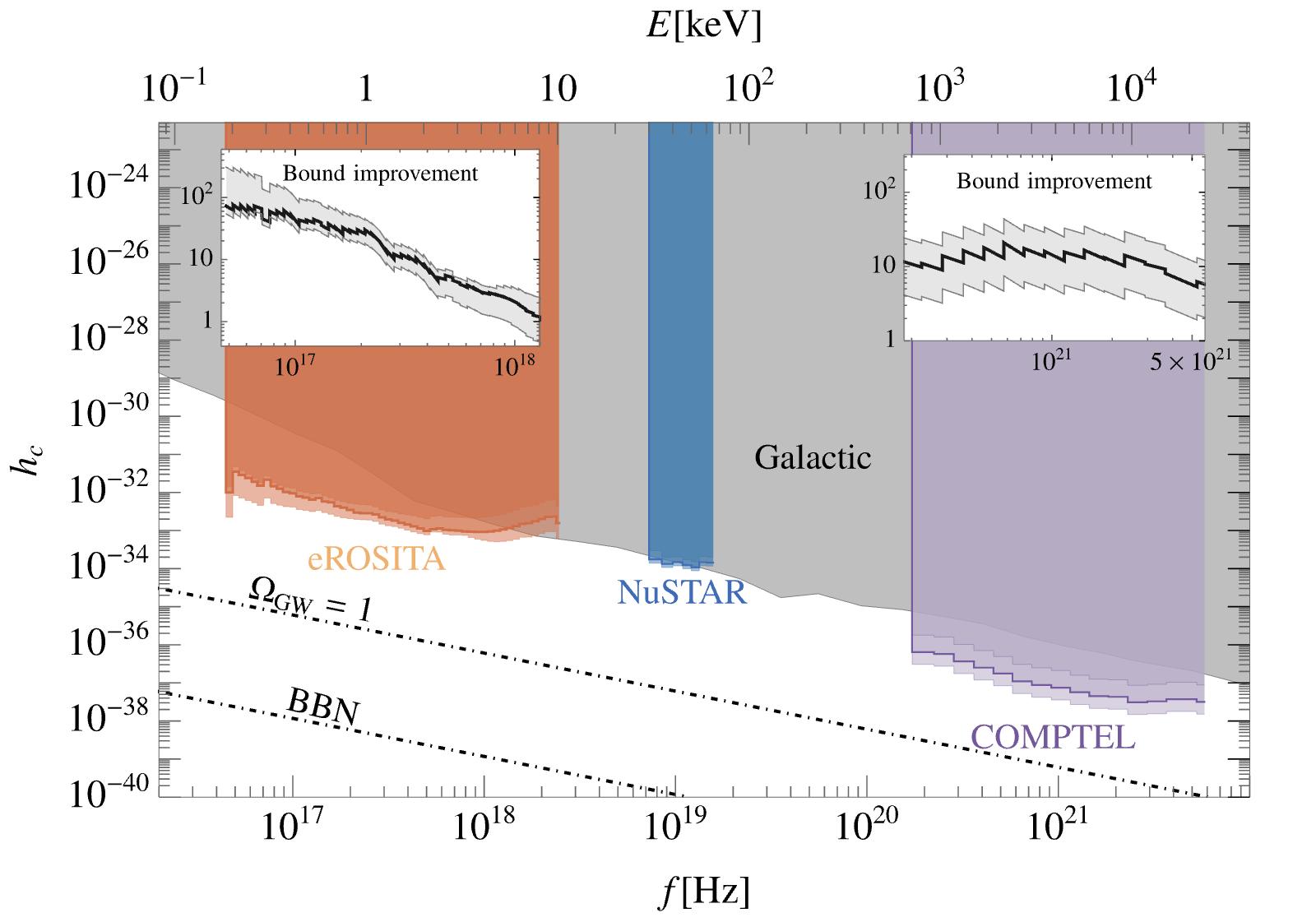}
    \caption{Exclusion limit bands on gravitational wave characteristic strain $h_c$ as a function of frequency $f$ (lower axis) and energy $E$ (upper axis). The orange area displays the bound obtained from eROSITA observations within an angular radius of $3.5^{\circ}$.} The blue area corresponds to the constraint derived from NuSTAR data as reported in Ref.~\cite{Ning:2024eky}, while the violet regions shows the region excluded by COMPTEL~\cite{Manna:2024tgu}. For comparison, we show limits from Ref.~\cite{Lella:2024dus}, based on graviton-to-photon conversions within the MW magnetic field. The subplot insets illustrate the factor of improvement of the $3.5^{\circ}$-FoV eROSITA and COMPTEL bounds relative to the Galactic bound from Ref.~\cite{Lella:2024dus}. Finally, the dot-dashed lines represent the BBN bound \cite{Pagano:2015hma} and the limit obtained by requiring that the gravitational wave energy density does not saturate the total energy density of the Universe.
    \label{fig:bound}
\end{figure*}

We evaluate the bounds from all $7$ sub-halos found in \textsc{IllustrisTNG}. For each frequency bin, we compute the minimum, median, and maximum bounds across the realizations.
The resulting bounds are illustrated in Fig.~\ref{fig:bound}, with the orange area corresponding to the eROSITA $3.5^\circ$ observations, while the blue and violet regions refer to the constraints derived from NuSTAR and COMPTEL observations. The minimum and maximum define the uncertainty bands, reflecting the systematic uncertainties from the magnetic field modeling, while the median is represented by the solid central line. We do not perform any likelihood analysis because uncertainties arising from the magnetic field modeling dominate over statistical fluctuations.
The insets in Fig.~\ref{fig:bound} display the factor of improvement achieved by our results compared to the Galactic conversion scenario analyzed in Ref.~\cite{Lella:2024dus}. Remarkably, in the frequency bands analyzed, our new bounds improve upon the Galactic constraint by 1–2 orders of magnitude.
Although the average magnitude of the magnetic field in the VC, $B_{\rm{VC}}\sim0.1\mu G$, is smaller by a factor of $3$ than in the MW, $B_{\rm MW}\sim 0.3\,\mu G$, the cluster environment provides a much longer baseline for graviton-to-photon conversions, $L_{\rm{VC}}\sim 1\,\rm{Mpc}$, compared to $L_{\rm{MW}}\sim 10\,\rm{kpc}$. Since the conversion probability in the coherent regime scales as $P_{g\gamma}\propto (BL)^2$, the fraction of gravitons converted while traversing the VC volume exceeds that in the MW by a factor of $\sim 10^3$.At lower frequencies $f\lesssim10^{18}\,{\rm Hz}$, where graviton-photon conversions are incoherent, this enhancement is even more pronounced. This is mainly due to the conservative estimate for $P_{g\gamma}$ employed in Ref.~\cite{Lella:2024dus}, which assumed a constant magnetic field along the LoS. As highlighted in Refs.~\cite{Carenza:2021alz,Marsh:2021ajy,Fiorillo:2026rgo}, magnetic field anisotropies on scales smaller than the size of the magnetic domain significantly magnify the conversion probability in the incoherent regime. The perturbative solution of the graviton-photon EoM developed in this work allows us to resolve the small-scale variations of the fields even in the frequency range where the typical oscillation length becomes smaller than such scales. As a result, it captures the stronger enhancement in the conversion probability~(see the Appendix~\ref{app:perturbative}) in the incoherent regime, which explains the greater enhancement observed in the frequency band probed by eROSITA compared to the COMPTEL energy band.

Despite the relatively large conversion probabilities, the photon flux that comes from VC and that reaches Earth suffers from the geometric dilution factor $1/D_{\rm VC}^2$ due to the large distance of the cluster $D_{\rm VC}$. This is where the use of large-FoV scanning survey instruments like eROSITA becomes crucial. As shown in Fig.~\ref{fig:FoV}, most of the observed X-ray emission from the VC is concentrated in the central region around M87, the brightest galaxy of the cluster. Nevertheless, the outer regions are expected to host non-negligible magnetic fields and can therefore provide relevant contributions to the HFGW conversion signal. Since these outer regions have much lower X-ray surface brightness than the VC core, increasing the extraction aperture enhances the integrated signal contribution without a proportional increase of the astrophysical X-ray background. As a result, the sensitivity improves for the larger extraction region, which is confirmed by our findings that eROSITA observations with a $3.5^\circ$ aperture yield substantially stronger constraints, outperforming the $0.5^\circ$ aperture by a factor $10$. For the same reason, VC observations conducted with small-FoV instruments that cover only the central region, such as NuSTAR with its $0.12^\circ$-radius FoV, result in comparatively weaker constraints.

This geometrical optimization of the aperture dictates how competing background noise affects our signal searches. Although a larger aperture maximizes the cluster signal, the resulting flux must be evaluated against the wider background environment.
To compare the amount of background affecting signal searches in the VC and the Galactic conversion scenarios, we recall that the isotropic electromagnetic flux produced by graviton conversions in the MW needs to be compared to the total cosmic flux from unresolved sources in the given frequency band. For instance, around $f\sim10^{17}\,{\rm Hz}$-- where we observe the largest improvement in our bounds-- the averaged Cosmic X-ray Background~(CXB) intensity inferred from ROSAT/ASCA measurements ~\cite{Miyaji:1998ez} corresponds to a photon flux of order $\sim 1\,{\rm photons\,cm^{-2}\,s^{-1}\,keV^{-1}}$ when integrated over the solid angle $3.5^\circ$ extraction region. The eROSITA spectra used in this work are not background-subtracted (see Figure~\ref{fig:spectra}) and should therefore be interpreted as the total observed emission within the aperture, blending any potential VC emission, with the CXB, Galactic foreground components, unresolved point sources, and instrumental/particle background. Taking the value at $f\sim10^{17}\,{\rm Hz}$ and dividing by the effective area of the instrument gives a flux of the order $\sim 3\,{\rm photons\,cm^{-2}\,s^{-1}\,keV^{-1}}$. Compared with the CXB-only estimate integrated over the same solid angle, this corresponds only to a modest enhancement of the total background level. Consequently, the enhanced conversion probability in the cluster environment directly translates into improved constraints with respect to the Galactic case.

Finally, our results need to be properly compared to the outcome of recent Ref.~\cite{Gupta:2026scx}, which used NuSTAR observations of the M87 galaxy in VC to obtain a bound at the level of $h_c\simeq1.2\times 10^{-36}$ at $10^{19}\,\rm{Hz}$, two orders of magnitude more stringent than our results.  A  side by side comparison using the same magnetic field model and observational parameters as in Ref.~\cite{Gupta:2026scx}, suggests that Gupta et al. likely neglect the geometrical penalty factor imposed by the limited FoV (discussed in Appendix~\ref{app:geometry}), which in the case of Ref.~\cite{EventHorizonTelescope:2024uoo} results in $45''$. When this limitation is properly taken into account, the resulting bounds are relaxed by several orders of magnitude.

\section{Conclusions}
\label{sec:conclu}

In this work, we have investigated the possibility of probing HFGWs through graviton–photon conversion in the galaxy cluster magnetic field, focusing on the VC as a representative astrophysical environment. By combining state-of-the-art magneto-hydrodynamical simulations from the \textsc{IllustrisTNG} project with multi-wavelength observations in the X-ray and soft $\gamma$-ray bands, we derived new constraints on the amplitude of a stochastic HFGW background. 

Our analysis demonstrates that galaxy clusters can act as efficient “cosmic lenses” for graviton–photon conversion. Although the magnetic field strength in the VC is typically lower than in the MW, the much larger spatial extension of the cluster significantly enhances the overall conversion probability in the coherent regime $P_{g\gamma} \sim  (BL)^2$, yielding an appreciable amplification of the induced photon flux over the Galactic case.

A key aspect of this work is the implementation of a perturbative treatment of graviton–photon mixing that accurately reproduces the full numerical solution while remaining computationally efficient. This approach allows us to account for the effects of  magnetic-field and electron-density fluctuations along individual LoSs through the cluster environment, resulting in an even stronger enhancement in the conversion probability in the incoherent oscillation regime compared to the coherent regime. 

Despite the large distance from the source, our constraints benefit from the use of large FoV instruments, able to cover at a glance a large portion of the Cluster. The resulting increase in the observed magnetic volume available for conversions is correlated to only a negligible increase of the stellar background from outer regions of the cluster, maximizing the signal-to-noise ratio for our exotic signals.

Using eROSITA and COMPTEL observations, we obtained competitive bounds on the characteristic strain $h_c$ in the frequency interval  $10^{16}$--$10^{22}$~Hz. The strongest improvements are achieved in the X-ray band probed by eROSITA, where graviton-photon oscillations are incoherent. In this frequency band our new constraints surpass previous Galactic limits by up to two orders of magnitude. In the coherent regime probed by COMPTEL, we find an improvement of approximately one order of magnitude relative to bounds derived from graviton–photon conversion in the MW magnetic field. 

Overall, our results highlight galaxy clusters as promising targets for indirect searches of HFGWs.
For comparison, Fig.~\ref{fig:bound} also shows the Big-Bang Nucleosynthesis (BBN) bound, which requires that the energy density stored in gravitational waves, $\rho_{\rm GW}$, does not exceed the cosmological energy density in relativistic species~\cite{Pagano:2015hma}. We stress, however, that the BBN constraint applies only to HFGWs produced before CMB decoupling, whereas our bounds are independent of the production epoch and therefore remain valid for any generation mechanism.  On the other hand, an unavoidable limit comes from requiring that GWs do
not saturate the total energy budget of the Universe.
This condition, displayed by the dot-dashed line labeled $\Omega_{\rm GW}=1$ in Fig.~\ref{fig:bound}, imposes a constraint on the characteristic strain that is approximately two orders of magnitude more stringent than the bounds derived in this work. Nevertheless, our limits offer a complementary probe to cosmological surveys in constraining HFGW backgrounds enhanced in non-standard cosmological scenarios, such as those generated by the evaporation of light PBHs~\cite{Dolgov:2013pwa, Ireland:2023avg}, which could in principle produce signals at the level of our sensitivities.

Looking ahead, future surveys in X-rays \cite{Gilli:2022vhs} and in $\gamma$-rays \cite{ DeAngelis:2021esn} are expected to further improve the sensitivity of these types of observations, confirming the high physics potential of the X-ray and $\gamma$-ray sky in constraining fundamental physics.

\acknowledgements

We warmly thank Maurizio Giannotti, Nicolao Fornengo, Marco Regis, Marco Taoso and Jorge Terol Calvo for the useful advices and fruitful discussions.
This article is based on work from COST Action COSMIC WISPers CA21106, supported by COST (European Cooperation in Science and Technology). 
AL is supported by the Italian MUR through the FIS 2 project FIS-2023-01577 (DD n. 23314 10-12-2024, CUP C53C24001460001), and by Istituto Nazionale di Fisica Nucleare (INFN) through the Theoretical Astroparticle Physics (TAsP) project.
JKV and FRC acknowledge support from the Deutsche Forschungsgemeinschaft (DFG, German Research Foundation) under Germany’s Excellence Strategy – Cluster of Excellence “Color meets Flavor”, EXC 3107 – Project-ID 533766364. JKV and FRC were also funded by the German federal and state program "Professorinnenprogramm 2030" Project-ID 01FP24167Q.
TZ acknowledge support from the Research grant TAsP (Theoretical Astroparticle Physics) funded by \textsc{infn}, from the European Union – Next Generation EU, from the Italian Ministry of University and Research (MUR) via the PRIN 2022 project n. 20228WHTYC - CUP D53C24003550006, and from the UniTO-IBS Avogadro Programme (collaboration agreement between the University of Torino and the Institute for Basic Science, South Korea).

\bibliographystyle{bibi}
\bibliography{references}

\setcounter{equation}{0}
\setcounter{figure}{0}
\setcounter{table}{0}
\setcounter{section}{0}
\makeatletter
\renewcommand{\theequation}{A\arabic{equation}}
\renewcommand{\thefigure}{A\arabic{figure}}
\renewcommand{\thetable}{A\arabic{table}}

\onecolumngrid

\clearpage

\appendix
\numberwithin{equation}{section}
\numberwithin{figure}{section}
\numberwithin{table}{section}

\section{Approximate expression for graviton-photon conversion probability}
\label{app:perturbative}
In this Appendix, we derive an approximate expression for the graviton--photon conversion probability. The derivation follows a perturbative approach similar to the treatment for Axion-like Particles (ALPs) described in Ref.~\cite{Evoli:2016zhj}. 
We recall that the relevant terms in the equation of propagation are shown in Eqs.~\eqref{eq:deltapl} and \eqref{eq:deltaggamma}.

Let us notice that in the graviton--photon oscillation the hierarchy $\Delta_{g\gamma}\ll \Delta_{\rm pl}$ is always satisfied in the environments taken into account. 
As shown in Sec.~\ref{sec:Graviton-photonconversion} the resulting EoM will be

\begin{equation}
i\frac{\partial}{\partial z}
\begin{pmatrix}
h_{+} \\ h_{\times} \\ A_x \\ A_y
\end{pmatrix}
=
\begin{pmatrix}
0 & 0 & \Delta_{g\gamma}\sin\phi & \Delta_{g\gamma}\cos\phi \\
0 & 0 & \Delta_{g\gamma}\cos\phi & -\Delta_{g\gamma}\sin\phi \\
\Delta_{g\gamma}\sin\phi & \Delta_{g\gamma}\cos\phi & \Delta_{\rm{pl}} & 0 \\
\Delta_{g\gamma}\cos\phi & -\Delta_{g\gamma}\sin\phi & 0 & \Delta_{\rm{pl}}
\end{pmatrix}
\begin{pmatrix}
h_{+} \\ h_{\times} \\ A_x \\ A_y
\end{pmatrix}.
\end{equation}

To solve this system, we perform a field redefinition
$
A_{x,y}(z)
=
\hat A_{x,y}(z)\,
\exp\!\left[-i\int_0^z ds\,\Delta_{\rm{pl}}(s)\right]
\equiv
\hat A_{x,y}(z)\,e^{-\alpha(z)} .
$
Since the mixing is weak $(\Delta_{g\gamma}\ll \Delta_{\rm{pl}})$ and the total conversion is small $(\Delta_{g\gamma}z\ll1)$, one can solve the equation of motion using the first order perturbation theory. For a magnetic field modeled as a sequence of $n$ coherent domains of length $l_p$ the photon field amplitudes at the end of the path are

\begin{align}
\hat A_x(z)
&=
-\frac{i}{\sqrt{2}M_{\rm Pl}}
\sum_{p=1}^{n}
\frac{B_{T,p}}{2}
\left(\sin\phi_p+\cos\phi_p\right)
\int_{z_p}^{z_{p+1}} d\eta\, e^{\alpha(\eta)},\\[6pt]
\hat A_y(z)
&=
-\frac{i}{\sqrt{2} M_{\rm Pl}}
\sum_{p=1}^{n}
\frac{B_{T,p}}{2}
\left(\cos\phi_p-\sin\phi_p\right)
\int_{z_p}^{z_{p+1}} d\eta\, e^{\alpha(\eta)},
\end{align}

where $\alpha(z_p)=i\int_{0}^{z_p}\Delta_{\rm{pl}}(s)ds$ is the accumulated phase and the integral over a single domain is 
\begin{equation}
    \int_{z_p}^{z_{p+1}} d\eta\, e^{\alpha(\eta)}
=
\ell_p\,
e^{\alpha(z_p)+i\Delta_{\rm{pl},p}\ell_p/2}\,
\mathrm{sinc}\!\left(\frac{\Delta_{\rm{pl},p} l_p}{2}\right).
\end{equation}

The total graviton--photon conversion probability is
\begin{equation}
P_{g\gamma}(z)=\frac{1}{2}\left(|\hat A_x(z)|^2 + |\hat A_y(z)|^2\right).
\end{equation}

\begin{figure}
    \centering
    \includegraphics[width=0.8\linewidth]{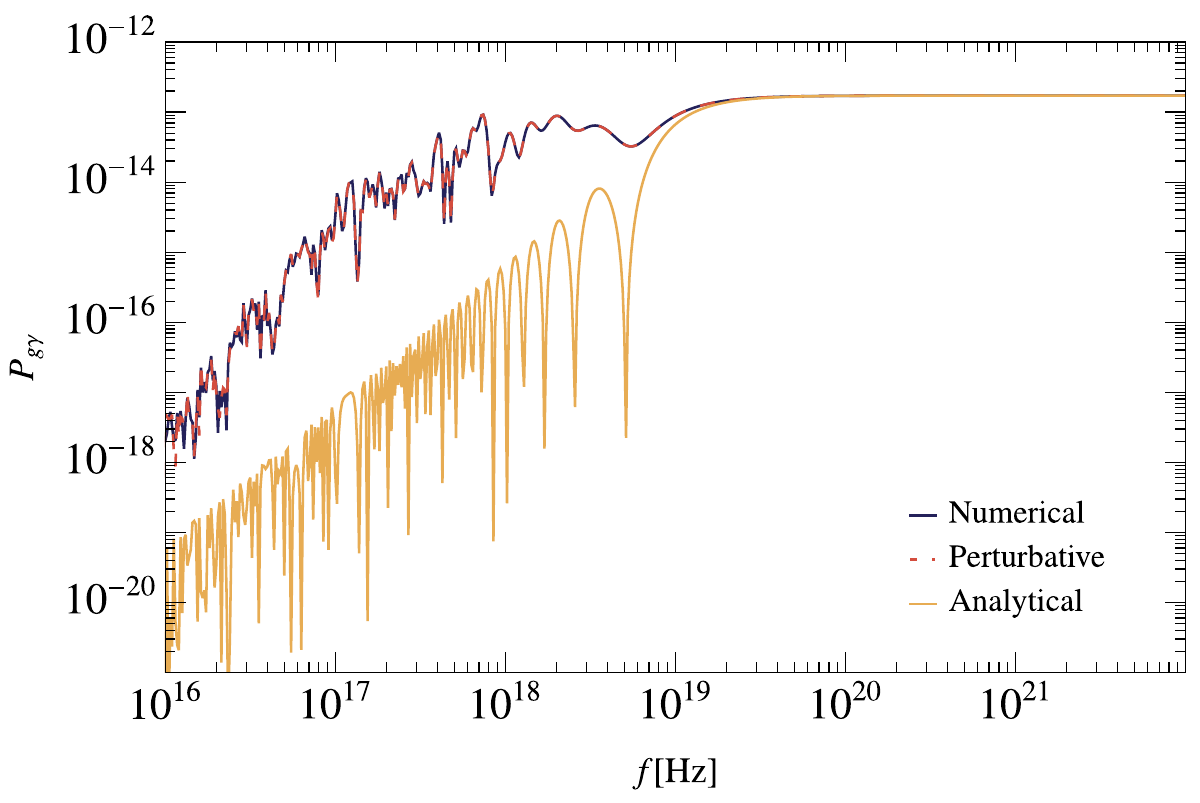}
    \caption{Graviton-photon conversion probabilities evaluated along a representative line of sight with three different approaches: solving the equation of motion (black curve), using the perturbative approach (red curve) and using the analytical formula for a homogeneous field (yellow curve).}
    \label{fig:plotA1}
\end{figure}

The accuracy of this perturbative approach is demonstrated in Fig.~\ref{fig:plotA1} , where we compare it against a full numerical integration of the EoM for a representative LoS. In particular, the probability conversion is presented for three distinct methods: the numerical solution of the EoM (black curve), the perturbative approach obtained in this Appendix (red dashed curve), and an approximation based on a homogeneous magnetic field (yellow curve). In the homogeneous case, the conversion probability is given by
\begin{equation}
    P_{g\gamma}=\frac{4\Delta^{2}_{g\gamma}}{\Delta_{\rm{osc}}^2}\sin^{2}\left(\frac{\Delta_{\rm{osc}}z}{2}\right)
\end{equation}
where the oscillation frequency is defined as
\begin{equation}
    \Delta_{\rm{osc}}=\sqrt{\Delta_{\rm pl}^2+4\Delta_{g\gamma}^2}
\end{equation}
For this analytical approximation, the probability is evaluated using the average transverse magnetic field component along the line of sight, defined as
\begin{equation}
    \langle B_T\rangle =\sqrt{ \frac{1}{L^2}\left(\left|\int_{0}^{L}B_x(z)dz\right|^2+\left|\int_{0}^{L}B_y(z)dz\right|^2\right)}
\end{equation}
and the average value of the electron density along the chosen line of sight
\begin{equation}
    \langle n_e\rangle = \frac{1}{L}\int_{0}^{L}n_e(z)dz\,.
\end{equation}
The agreement between the numerical and perturbative results is excellent across the entire energy range. Notably, the perturbative approach offers a significant gain in computational efficiency, reducing the calculation time to $\mathcal{O}(10^{-2})$ s per line of sight.

\section{Impact of the hard edges}
\label{app:hardedges}

\begin{figure}[t!]
    \centering
    \includegraphics[width=0.8\linewidth]{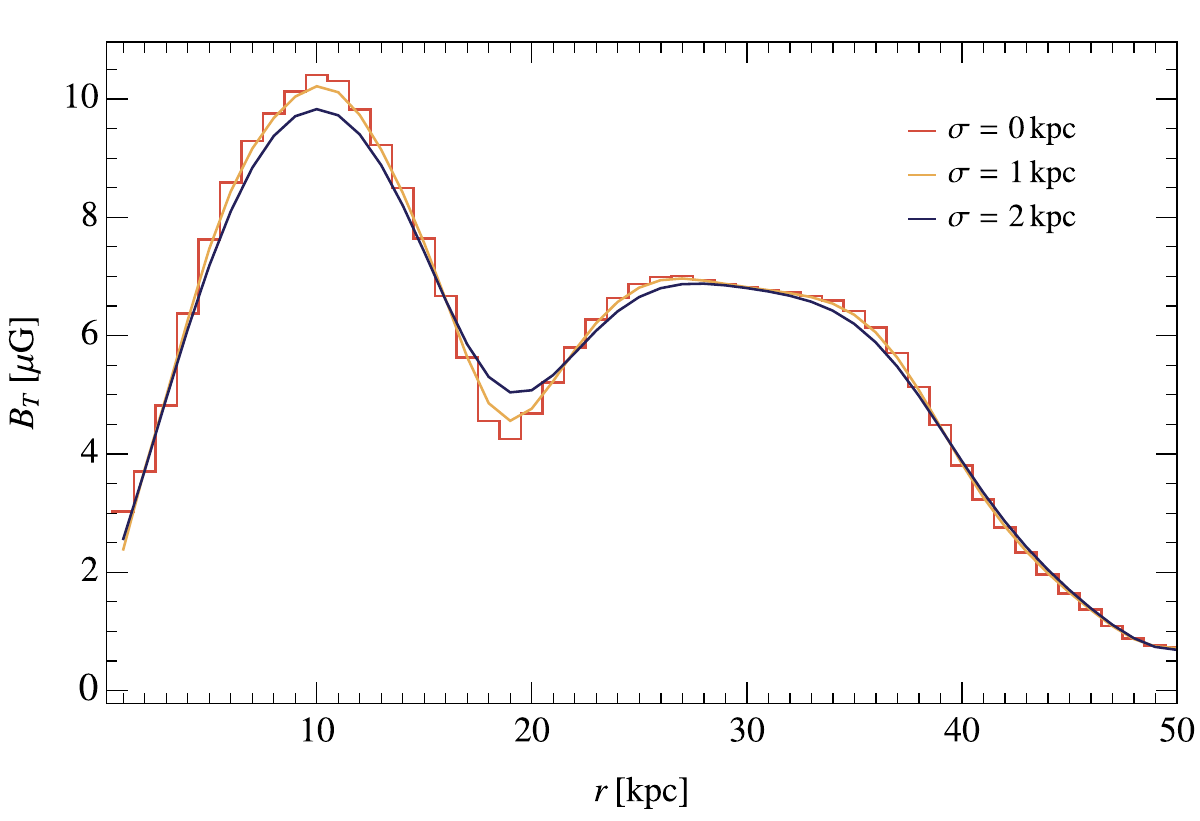}
    \caption{Zoomed-in representation of the magnetic field of a LoS of the VC as used in the perturbative approach (red), alongside curves obtained by smoothing the magnetic field with $\sigma = 1\,\text{kpc}$ (yellow) and $\sigma = 2\,\text{kpc}$ (blue).}
    \label{fig:BTsmoothplot}
\end{figure}

\begin{figure}[t!]
    \centering
    \includegraphics[width=0.8\linewidth]{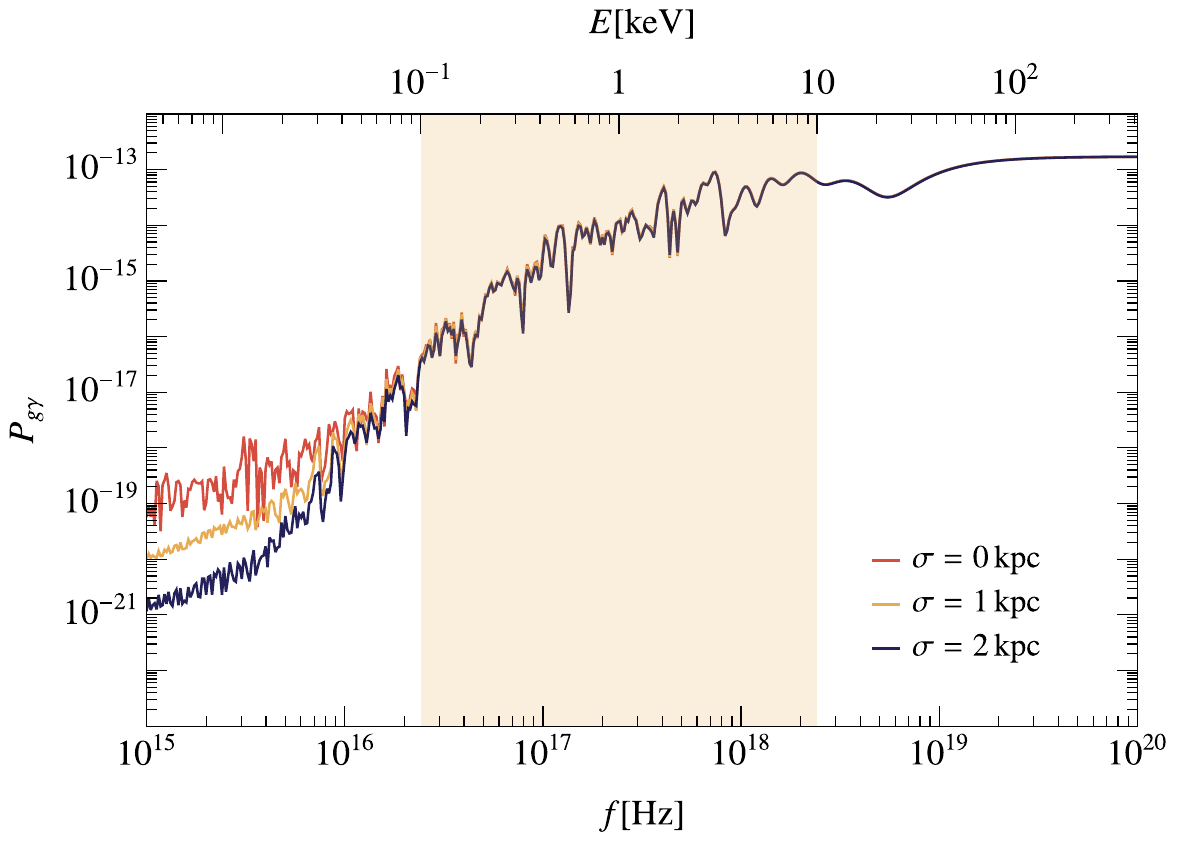}
    \caption{Graviton-to-photon conversion probabilities associated with the three magnetic field configurations shown in the top panel. The orange shaded band indicates the frequency range in which eROSITA is operational.}
    \label{fig:Pgsmoothplot}
\end{figure}

A potential issue in evaluating the graviton-to-photon conversion using finite-resolution magneto-hydrodynamical simulations (such as the profile shown in Fig.~\ref{fig:plotneBTcomb}) is the presence of sharp cell boundaries. As discussed in the previous literature (Refs.~\cite{Marsh:2021ajy,Galanti:2018nvl}), hard-edge cell boundaries enhance the magnetic field power spectrum at small scales, leading to an overestimation of the conversion probabilities.
To test the robustness of our results against this effect, we examine the stability of the conversion probabilities under Gaussian smoothing, following an approach analogous to that in Ref.~\cite{Lella:2024dus}. Gaussian smoothing removes these nonphysical hard-edge gradients while preserving the morphology of the cluster magnetic field. Specifically, we smooth the magnetic field along a representative LoS using smoothing lengths of $\sigma =1\,\rm{kpc}$ and $\sigma=2\,\rm{kpc}$. The smoothed field profile can be then obtained as
\begin{equation}
    B_{\rm smooth}(x)=\int dy \frac{1}{\sqrt{2\pi \sigma^2}} e^{-\frac{(x-y)^2}{2\sigma}} B_T(y)\,.
\end{equation}
At this point, the transition amplitude for graviton-photon conversion will be
\begin{equation}
   \mathcal{T}_{i}(z)\simeq -\frac{i}{\sqrt{2}M_{\rm pl}} \int_{z_0}^z dz'\,\Delta_{g\gamma,i}(z')\,e^{\alpha(z')}
\end{equation}
where the index $i=x,y$ corresponds to the two polarization components with $\Delta_{g\gamma,x}=\Delta_{g\gamma}\sin\phi$ and $\Delta_{g\gamma,y}=\Delta_{g\gamma}\cos\phi$ and $\alpha(z')=i\int_{0}^{z'}\Delta_{\rm pl}(s)ds$ represents the accumulated plasma phase.
A zoomed-in view of a LoS of the VC magnetic field configuration and the effect of smoothing at different scales are shown in Fig.~\ref{fig:BTsmoothplot}, while the resulting graviton-to-photon conversion probabilities are shown in Fig.~\ref{fig:Pgsmoothplot}. Any noticeable discrepancy between the profiles becomes relevant below $10^{16}\,\rm{Hz}$, which falls outside our frequency range of interest. Indeed, at $10^{16}$ Hz the oscillation length $l_{\rm{osc}}=2\pi/\Delta_{\rm{osc}}$ becomes comparable to the correlation length of the field $\sim 10\,\rm{kpc}$, meaning the conversion is strongly affected by the magnetic field structure over small scales.

\section{Flux integration over the cluster volume}
\label{app:geometry}
\begin{figure}
    \centering
    \includegraphics[width=0.7\linewidth]{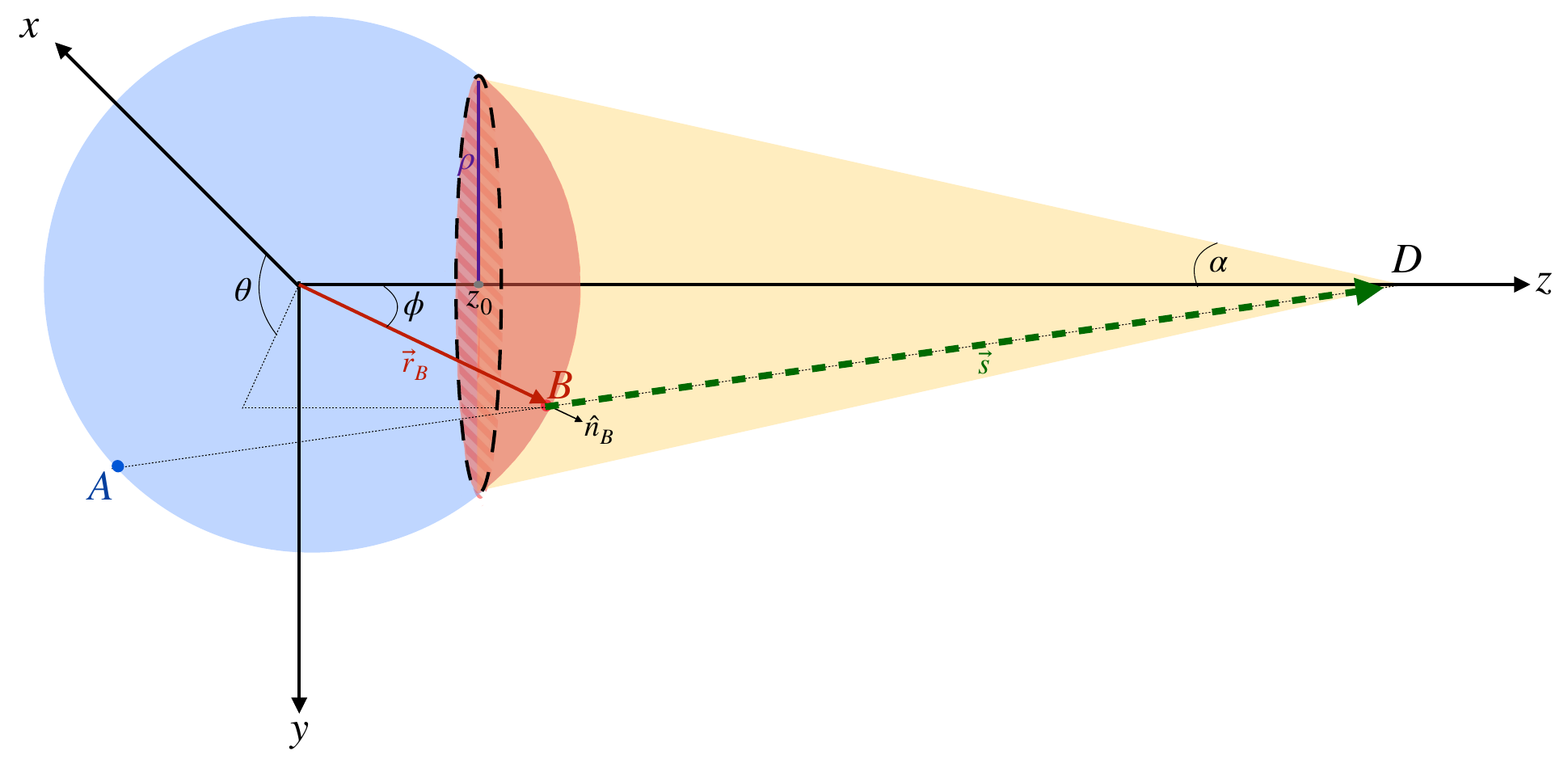}
    \caption{Geometrical setup of the graviton-to-photon conversion model. The host environment is modeled as a sphere of radius $R$ (blue sphere) centered at the origin of the Cartesian coordinate system, with an observer positioned at distance $D$ along the $z$-axis (gray point). The instrument's FoV is represented by a cone (yellow cone) with a half-opening angle $\alpha$. The intersection of the cone and the sphere defines the visible volume (red spherical cap) with a base located at height $z_0$ (gray point) and a transverse radius $\rho$ (purple segment). For a representative line of sight point $A$ (blue point) identifies the entry of the graviton flux into the host environment magnetic field, while point $B$ (red point) marks the exit point toward the observer. Point $B$ is identified by a position vector $\vec{r}_B$ (red vector), which defines the outward unit normal to the surface $\hat{n}_B$ (black vector). The displacement vector connecting the exit point $B$ and the observed is denoted by $\vec{s}$ (dashed green vector). The spherical coordinated $\theta$ and $\phi$ used for the integration and surface parametrization are also indicated.}
    \label{fig:plotA2}
\end{figure}

In this Appendix, we provide a derivation of the geometric model used to calculate the integrated photon flux produced by graviton-to-photon conversions. The aim is to determine the volume of the host environment visible within the instrument's FoV and define the integration element for the observed signal. The geometric configuration is illustrated in Fig.~\ref{fig:plotA2}.
We model the host environment as a sphere of radius $R$, centered in the origin of the cartesian coordinate system $(x,y,z)$ (blue sphere, Fig.~\ref{fig:plotA2}). The observer is located at a distance $D$ from the center and is positioned, without loss of generality, along the positive $z$-axis at position 
\begin{equation}
    \vec{r}_{\rm{obs}}=(0,0,D)\,.
\end{equation} 
The instrument's FoV is modeled as a cone with its apex at the observer's position, oriented toward the origin (yellow cone, Fig.~\ref{fig:plotA2}). The cone is characterized by a half-opening angle $\alpha$.
The region where graviton-to-photon conversions contribute to the observed signal is defined by the intersection of the instrument's FoV with the spherical host environment. The boundary of this visible region is determined by the intersection of
\begin{align}
x^2 + y^2 + z^2 &= R^2 \label{eq:sphere_app} \\
\sqrt{x^2 + y^2} &= (D - z)\tan\alpha \quad \label{eq:cone_app}
\end{align}
This intersection defines a circle perpendicular to the $z$-axis at a coordinate $z_0$ (gray point, Fig.~\ref{fig:plotA2}). By substituting Eq.~\eqref{eq:cone_app} into Eq.~\eqref{eq:sphere_app}, we find the following quadratic equation and rearranging the terms we obtain
\begin{equation}
z^2(1 + \tan^2\alpha) - 2Dz\tan^2\alpha + (D^2\tan^2\alpha- R^2) = 0.
\end{equation}
The discriminant of this quadratic equation is proportional to $R^2-D^2 \sin^2\alpha$. Real solutions for the intersection exist only when the condition $R\geq D\sin\alpha$ is satisfied. When $R<D\sin\alpha$ (the large FoV regime), the discriminant is negative and the cone does not physically intersect the sphere. The visible region is instead bounded by LoSs tangent to the sphere.
For the regime where real solutions exist, the solution $z_0$ corresponds to the $z$-coordinate of the boundary of the spherical cap visible to the observer (red cap, Fig.~\ref{fig:plotA2}). The transverse radius of this intersection is then given by $\rho=\sqrt{R^2-z_0^2}$ (purple segment, Fig.~\ref{fig:plotA2}).
The total converted photon flux is determined by integrating the conversion probability $P_{g\gamma}$ over the LoSs that span the FoV. 
\begin{equation}
\frac{dF_{g}}{df} = \int_{\rm vis.\,surf.} d\Omega \; \frac{d^2F_{g}}{df d\Omega}P_{g\gamma}(f,\Omega)
\end{equation}
where $\frac{d^2F_{g}}{df d\Omega}$ denotes the flux shown in Eq.~\eqref{eq:gravitonflux} and $d\Omega$ is the solid-angle element subtended by the visible surface. 
An explicit form of the differential solid angle can be obtained by defining the flux exit point B (red point in Fig.~\ref{fig:plotA2}), which is identified by the position vector
\begin{equation}
    \vec{r}_B=R(\sin \phi \cos \theta, \sin \phi \sin \theta, \cos \phi)\,.
\end{equation}
From this, the corresponding outward unit normal is defined as $\hat{n}_B=\vec{r}_B/R$ (black vector in Fig.~\ref{fig:plotA2}).
The vector that connects the emission point $B$ to the observer is then given by the difference 
\begin{equation}
    \vec{s}=\vec{r}_{\rm obs}-\vec{r}_B=(-R\sin \phi \cos \theta,-R\sin \phi \sin \theta ,D-R\cos \phi)
\end{equation}
shown as a green dashed arrow in Fig.~\ref{fig:plotA2}.
Consequently, the differential solid angle is expressed as
\begin{equation}
    d\Omega=\frac{\hat{n}_B\cdot \vec{s}}{||\vec{s}||^3} dA=\frac{D\mu-R}{(D^2+R^2-2DR\mu)^{3/2}}
R^2 d\mu d\theta 
\label{eq:solidangle}
\end{equation}
where $\mu=\cos \phi$. 

Based on the defined geometry, the $z$ coordinate is given by $z=R\cos\phi=R\mu$. The lower limit of integration is determined by the intersection of the cone and the sphere, $\mu_{\rm{min}}=z_0/R$ , or by the tangency condition $\mu_{\rm{min}}=R/D$ in the large-aperture regime where the intersection fails. The integration domain is therefore $\theta\in [0,2\pi)$ and $\mu\in[\mu_{\rm min},1]$. Performing the integration of Eq.~\eqref{eq:solidangle} over this region yields the total solid angle
\begin{equation}
    \Omega=2\pi\left(1-\frac{D-R\mu_{\rm{min}}}{\sqrt{D^2+R^2-2DR\mu_{\rm{min}}}}\right)
    \,\ .\label{eq:intsolidangle}
\end{equation}

While the analytical expression for the solid angle shown in Eq.~\eqref{eq:intsolidangle} provides a useful geometric estimate of the visible region, the total photon flux is computed numerically by explicitly integrating along discrete LoS within the FoV. The visible spherical cap has been discretized and each sampled point has been treated as an exit point $B_i$ (indicated as a red point in Fig.~\ref{fig:plotA2}), where the converted photon leaves the host environment toward the observer.
For each exit point $B_i$, the LoS is defined as the segment connecting the observer to that point. The path is parameterized as,
\begin{equation}
\vec{P}(t) = \vec{r}_{\rm obs} + t (\vec{r}_{B_i} - \vec{r}_{\rm obs}),
\end{equation}
with $t=0$ at the observer location and $t=1$ at the exit point $B_i$. The corresponding entry point of the graviton flux in the host magnetic field $A_i$ (indicated as a blue point in Fig.~\ref{fig:plotA2}) is determined by finding the intersection of this line with the spherical boundary. This is achieved by solving for $t$ the following equation
\begin{equation}
||\vec{r}_{\rm obs}||^2 + 2t  [\vec{r}_{\rm obs} \cdot (\vec{r}_{B_i} - \vec{r}_{\rm obs})] + t^2 ||\vec{r}_{B_i} - \vec{r}_{\rm obs}||^2 = R^2.
\end{equation}
The solution to this equation identifies the entry point $A_i$ for a fixed exit point $B_i$. For each sampled LoS, the relevant physical quantities are evaluated along the segment between $A_i$ and $B_i$, and the conversion probability $P_{g\gamma}$ is evaluated by integrating along this path. Finally, each contribution is weighted by the differential solid angle element given in Eq.~\eqref{eq:solidangle}, and the total flux is obtained by summing over all sampled lines of sight within the FoV.

\section{Dependence on Virgo Cluster Modeling}
\label{app:Marsh}
\begin{figure*}
    \centering
    \includegraphics[width=0.8\linewidth]{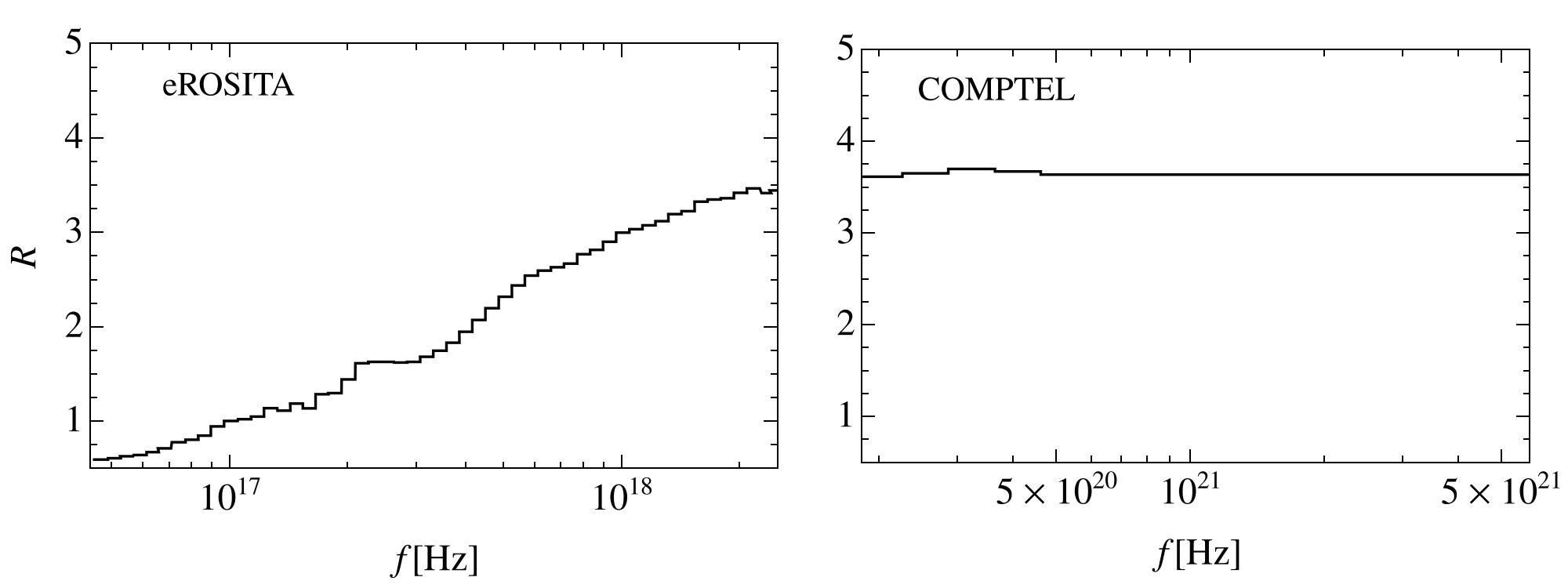}
    \caption{Ratio between the median bounds $R$ obtained using magnetic field realizations based on the RF model of Ref.~\cite{Marsh:2017yvc} and those obtained from the \textsc{IllustrisTNG} realizations. The left panel shows the ratio in the eROSITA frequency range, while the right panel shows the corresponding ratio in COMPTEL range.}
    \label{fig:boundwMarsh}
\end{figure*}

To assess the robustness of the results against the modeling of the VC magnetic field, we compare the constraints obtained using the \textsc{IllustrisTNG} simulations with those derived from the magnetic-field model of Ref.~\cite{Marsh:2017yvc}. In that framework, the magnetic field is formulated as a 1D radial profile with an orientation that varies randomly from cell to cell. The correlation length, $l_{\rm corr}$, is sampled from the following distribution
\begin{equation}
      p(l_\mathrm{corr})=
      \begin{cases}
      l_\mathrm{corr}^{-\gamma}\,\,\,\,\mathrm{for} \,l_\mathrm{corr}>l_{\rm min} \\
      0\,\,\,\,\mathrm{otherwise}
      \end{cases}
\end{equation}
with a shape parameter is $\gamma=2.5$. For radial distances of $|r|\leq 2\,\mathrm{kpc}$, the minimum correlation lenght is set to $l_{\mathrm{min}}=0.2\,\mathrm{kpc}$. 
\begin{figure}
    \centering
    \includegraphics[width=0.8\linewidth]{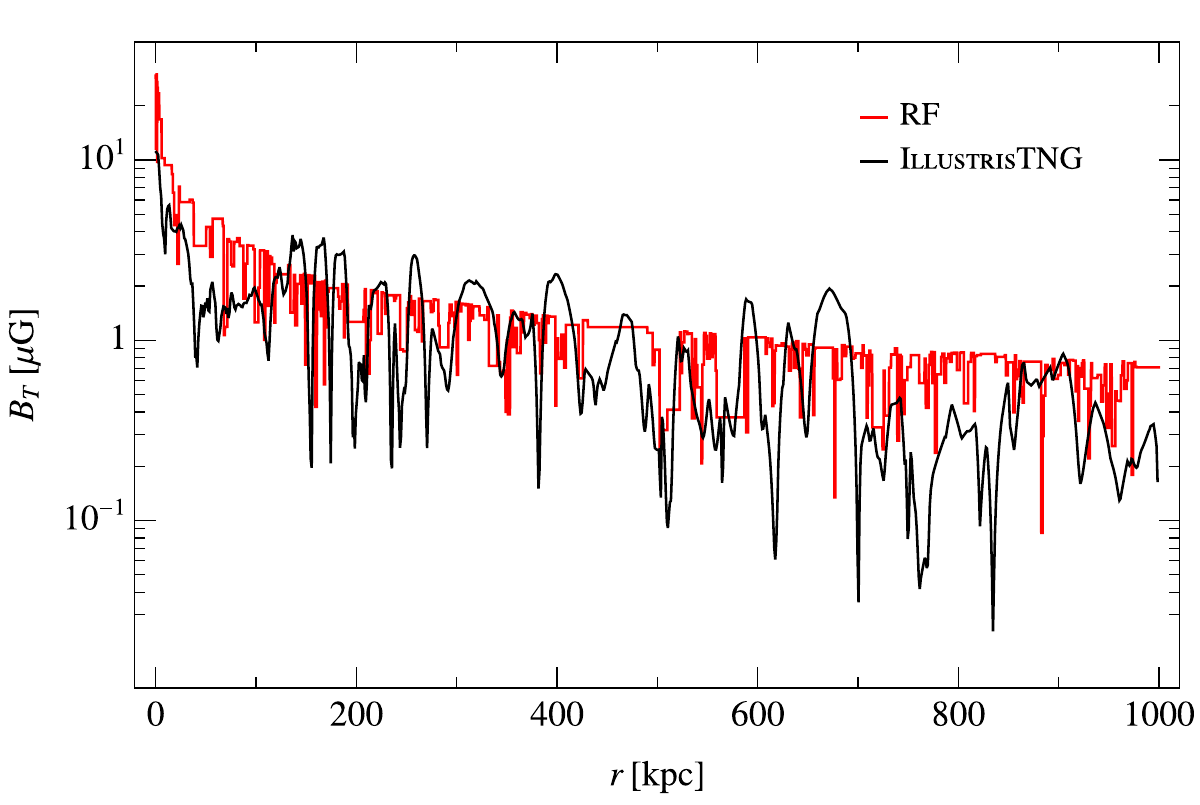}
    \caption{Radial profile of the magnetic field of the VC from the center out to $1\,\rm{Mpc}$. The black curve shows the profile for a representative LoS of an \textsc{IllustrisTNG} subhalo, while the red curve represents a LoS from the center of a realization based  on the random field (RF) model of Ref.~\cite{Marsh:2017yvc}.
    }
    \label{fig:BTradialMarsh}
\end{figure}
To perform a consistent comparison with our analysis, we extended this 1D prescription to construct $10^2$ full 3D magnetic-field configurations. Specifically, we built each 3D realization by generating independent 1D radial profiles along different angular directions $(\theta,\phi)$ outward from the cluster center, effectively assembling these independent radial LoSs into a complete spherical geometry. 
For each realization, we evaluated the graviton-to-photon conversion probability and derived the associated bounds. We then compared the median constraints obtained from these realizations with those obtained using the \textsc{IllustrisTNG} magnetic-field configurations.
While both models yield comparable average transverse magnetic-field strengths, they differ substantially in their spatial structure. Notably, the characteristic correlation length in the random field (RF) model of Ref.~\cite{Marsh:2017yvc} is roughly a factor of three smaller than that found in the \textsc{IllustrisTNG} realizations. Fig.~\ref{fig:BTradialMarsh} displays the magnetic field profile of the VC from the center out to $1\,\rm{Mpc}$, comparing a central LoS profile of an \textsc{IllustrisTNG} subhalo (black) with a central LoS from a realization based on the RF model of Ref.~\cite{Marsh:2017yvc} (red). 
The impact of these differences on the final constraints is shown in Fig.~\ref{fig:boundwMarsh}, where we show the ratio of the bounds on $h_c$ obtained using the model of Ref.~\cite{Marsh:2017yvc} to the ones obtained with the \textsc{IllustrisTNG} realizations,
\begin{equation}
R(f) \equiv
\frac{h_c^{\mathrm{RF}}(f)}
     {h_c^{\mathrm{TNG}}(f)} \,\ , 
\end{equation}
A ratio $R(f)>1$ indicates that the
\textsc{IllustrisTNG} configurations yield stronger constraints,
whereas $R(f)<1$ implies stronger constraints from the RF model of
Ref.~\cite{Marsh:2017yvc}.
In the COMPTEL frequency range, the ratio is $R \sim 3$ and remains nearly constant. Indeed, the conversion probabilities obtained with the realizations from the RF model of Ref.~\cite{Marsh:2017yvc} are about one order of magnitude smaller than those obtained with the \textsc{IllustrisTNG} configurations. Since the bound scales as $h_c\propto P_{g\rightarrow\gamma}^{-1/2}$, this translates into constraints weaker by a factor of about $3$, which is consistent with the value observed in Fig.~\ref{fig:boundwMarsh}. The constant behavior arises because the graviton-to-photon conversion probability has reached the coherent regime at COMPTEL frequencies, rendering the relative difference between the two magnetic field models essentially frequency-independent.  
A different behavior emerges in the eROSITA range. Here, for frequencies $f\lesssim 10^{17}\rm{Hz}$, the bounds based on the RF model of Ref.~\cite{Marsh:2017yvc} model are comparable to, or even stronger than, those from \textsc{IllustrisTNG}. At higher frequencies, the ratio progressively increases to $R \sim 3$, meaning the \textsc{IllustrisTNG} realizations provide stronger constraints in the upper eROSITA range. 
This trend can be interpreted in terms of the phase coherence of the conversion amplitude within a magnetic domain. Contributions generated at different positions add constructively only if their relative phase remains small. Thus, even within a region where the magnetic field maintains approximately the same orientation, a sufficiently large phase variation can lead to partial cancellation of the conversion amplitude. Across a domain of size $l_{\rm corr}$, the accumulated phase is approximately
\begin{equation}
\delta\alpha \sim |\Delta_{\rm pl}|l_{\rm corr}
= \frac{2\pi l_{\rm corr}}{l_{\rm osc}},
\end{equation}
where $l_{\rm osc}\simeq 2\pi/|\Delta_{\rm pl}|$ in the plasma-dominated regime relevant to this discussion. Coherent accumulation within a domain requires $\delta\alpha\ll 1$, or equivalently $l_{\rm osc}\gg 2\pi l_{\rm corr}$. Since $|\Delta_{\rm pl}|\propto n_e/f$, increasing the frequency reduces the phase variation across a given domain and therefore tends to alleviate phase-induced cancellations.
At $f=5\times 10^{16} \,\ \mathrm{Hz}$, the average electron densities of the realizations give characteristic oscillation lengths of $l_{\rm osc}\sim 8 \,\  \mathrm{kpc}$ for the RF model of Ref.~\cite{Marsh:2017yvc} and $\sim 5\,\mathrm{kpc}$ for \textsc{IllustrisTNG}. The corresponding magnetic correlation lengths are $l_{\rm corr}\sim 3 \,\  \mathrm{kpc}$ and $\sim 10 \,\ \mathrm{kpc}$, respectively. Neither phase variation  lies in the fully coherent regime, but the substantially larger phase variation in the \textsc{IllustrisTNG} configurations suggests that cancellations within correlated regions can play a more prominent role. This explains why their larger field correlation length does not necessarily yield stronger constraints at the lower end of the eROSITA frequency range.
At $f\sim 10^{17}\,\mathrm{Hz}$, the estimated oscillation lengths increase to approximately $16 \,\ \mathrm{kpc}$ and $10 \,\mathrm{kpc}$, reducing the corresponding phase variations. Although the \textsc{IllustrisTNG} domains are still not fully coherent at this frequency, the decreasing phase variation is consistent with a gradual reduction of phase-induced suppression as the frequency increases. Once this suppression becomes sufficiently weak, more extended correlated regions can favor the accumulation of the conversion amplitude, providing a qualitative interpretation of the stronger \textsc{IllustrisTNG} constraints at higher frequencies. The precise frequency dependence and the magnitude of the ratio nevertheless depend on the full magnetic-field and electron-density profiles and must be obtained from the conversion probabilities computed for each realization.

\end{document}